\documentclass[preprint,3p,times, nonatbib, twocolumn]{elsarticle}

\usepackage{tikz}
\usetikzlibrary{shapes.geometric}
\usepackage{amssymb}
\usepackage{xspace}
\usepackage[version=4]{mhchem}
\usepackage[font=footnotesize,labelfont=bf,skip=5pt,justification=RaggedRight]{caption}
\usepackage{subcaption}
\usepackage{multirow}
\usepackage{amsmath}
\usepackage{slashed}
\usepackage{braket}
\usepackage{booktabs}
\usepackage{tabularx}
\usepackage{placeins}
\usepackage{xurl}
\usepackage{hyperref}
\hypersetup{
    colorlinks=true,
    linkcolor=blue,
    filecolor=magenta,      
    urlcolor=cyan,
    pdftitle={SET-ANUBIS: a modular pipeline for ANUBIS long-lived particle sensitivity studies},
    pdfpagemode=FullScreen,
    }
\usepackage[most]{tcolorbox}
\usepackage{listings}
\usepackage{xcolor}

\definecolor{mykeywords}{RGB}{15,135,70}
\definecolor{mygray}{RGB}{250,250,250}
\definecolor{mystrings}{RGB}{50,110,100}
\definecolor{mycomments}{RGB}{150,200,205}

\usepackage{lineno}

\newcounter{bla}

\journal{Computer Physics Communications}

\newcommand{\setanubisrepo}{\url{https://github.com/SET-ANUBIS/set-anubis}}
\newcommand{\setanubisdocs}{\url{https://set-anubis.github.io/set-anubis/}}
\newcommand{\setanubistag}{\texttt{v1.0.0}\xspace}
\newcommand{\setanubisdoi}{\url{https://doi.org/10.5281/zenodo.21462101}}
\newcommand{\setanubispypi}{\texttt{SetAnubis==1.0.0}\xspace}
\newcommand{\setanubispypiurl}{\url{https://pypi.org/project/SetAnubis/}}
\newcommand{\setanubistestpypiurl}{\url{https://test.pypi.org/project/SetAnubis/}}

\begin{document}
\newcommand{\quotes}[1]{``#1''}
\newcommand{\eg}{\mbox{\itshape e.g.}\xspace}
\newcommand{\ie}{\mbox{\itshape i.e.}\xspace}
\newcommand{\etal}{\mbox{\itshape et al.}\xspace}
\newcommand{\etc}{\mbox{\itshape etc.}\xspace}
\newcommand{\cf}{\mbox{\itshape cf.}\xspace}
\newcommand{\ffp}{\mbox{\itshape ff.}\xspace}
\newcommand{\vs}{\mbox{\itshape vs.}\xspace}

\newcommand{\proanubis}{\mbox{\textsl{pro}ANUBIS}\xspace}
\newcommand{\anubis}{\mbox{ANUBIS}\xspace}
\newcommand{\setanubis}{\mbox{\textsc{SET-ANUBIS}}\xspace}
\newcommand{\osiris}{\mbox{\textsc{OSIRIS}}\xspace}
\newcommand{\atlas}{\mbox{ATLAS}\xspace}

\newcommand{\pythia}[1][]{\mbox{\textsc{Pythia#1}}\xspace}
\newcommand{\madgraph}{\mbox{\textsc{MadGraph}}\xspace}
\newcommand{\madspin}{\mbox{\textsc{MadSpin}}\xspace}
\newcommand{\madversion}[1]{MG5\_aMC@NLO #1}
\newcommand{\marty}{\mbox{\textsc{MARTY}}\xspace}
\newcommand{\mathematica}{\mbox{\textsc{Mathematica}}\xspace}
\newcommand{\powheg}{\mbox{\textsc{Powheg}}\xspace}
\newcommand{\geant}{\mbox{\textsc{Geant4}}\xspace}

\newcommand{\ns}{\text{ns}\xspace}
\newcommand{\ps}{\text{ps}\xspace}
\newcommand{\mum}{\ensuremath{\mu\textnormal{m}}\xspace}
\newcommand{\mm}{\ensuremath{\textnormal{mm}}\xspace}
\newcommand{\cm}{\ensuremath{\textnormal{cm}}\xspace}
\newcommand{\metre}{\ensuremath{\textnormal{m}}\xspace}
\let\meter=\metre
\newcommand{\MeV}{\ensuremath{\textnormal{MeV}}\xspace}
\newcommand{\GeV}{\ensuremath{\textnormal{GeV}}\xspace}
\newcommand{\TeV}{\ensuremath{\textnormal{TeV}}\xspace}
\newcommand{\rad}{\ensuremath{\textnormal{rad}}\xspace}
\newcommand{\mrad}{\ensuremath{\textnormal{mrad}}\xspace}
\newcommand{\pb}{\ensuremath{\rm pb}\xspace}
\newcommand{\fb}{\ensuremath{\rm fb}\xspace}
\newcommand{\ab}{\ensuremath{\rm ab}\xspace}
\newcommand{\invpb}{\ensuremath{{\rm pb}^{-1}}\xspace}
\let\ipb=\invpb
\newcommand{\invfb}{\ensuremath{{\rm fb}^{-1}}\xspace}
\let\ifb=\invfb
\newcommand{\invab}{\ensuremath{{\rm ab}^{-1}}\xspace}
\let\iab=\invab
\newcommand{\stat}{\ensuremath{\textnormal{(stat)}}\xspace}
\newcommand{\syst}{\ensuremath{\textnormal{(syst)}}\xspace}

\newcommand{\et}{\ensuremath{E_{\mathrm{T}}}\xspace} 
\newcommand{\met}{\ensuremath{E_T^\text{miss}}\xspace} 
\newcommand{\pt}{\ensuremath{p_T}\xspace} 
\newcommand{\order}[1]{\ensuremath{\mathcal{O}(#1)}\xspace} 
\newcommand{\leff}{\ensuremath{L_\text{eff}}\xspace} 
\newcommand{\const}{\operatorname{const}}
\newcommand{\eps}{\varepsilon} 
\newcommand{\acc}{\ensuremath{\mathcal A}\xspace} 
\def\sqm{\ensuremath{\text{m}^2}\xspace} 
\def\mllp{\ensuremath{m_\text{LLP}}\xspace} 
\def\meff{\ensuremath{m_\text{eff}}\xspace} 
\def\mmed{\ensuremath{m_\text{med}}\xspace} 
\newcommand{\dif}{\ensuremath{{\rm d}}\xspace}
\newcommand{\del}{\partial}
\newcommand{\dR}{\ensuremath{\Delta R}\xspace}
\def\dtheta{\ensuremath{\Delta\theta}\xspace} 
\def\dphi{\ensuremath{\Delta\phi}\xspace} 
\newcommand{\brx}[1]{\ensuremath{\mathcal Br_{#1}}\xspace}
\newcommand{\br}{\ensuremath{\mathcal Br}\xspace}
\let\BF=\br
\newcommand{\hepgroup}[2]{\ensuremath{\mathbb{#1}_{#2}}\xspace}
\newcommand{\SU}{\ensuremath{\mathbb{SU}}\xspace} 
\newcommand{\U}{\ensuremath{\mathbb{U}}\xspace} 
\newcommand{\LI}{\ensuremath{\Lambda_I}\xspace}
\newcommand{\LIx}[1]{\ensuremath{\Lambda_{I,\text{#1}}}\xspace}

\begin{frontmatter}




\title{{\LARGE \texorpdfstring{SET-ANUBIS: a modular pipeline to perform long-lived particle sensitivity studies for the ANUBIS detector\\ 
\includegraphics[width=0.1\linewidth]{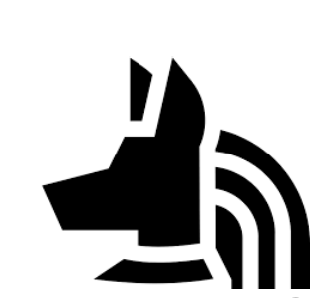}\\
{\large\textbf{ANUBIS Collaboration}}
\\[-0.2em]
{\normalsize\textit{E-mail: }\href{mailto:anubis-publications@cern.ch}{anubis-publications@cern.ch}}
\\[-0.2em]
}{SET-ANUBIS: a modular pipeline for ANUBIS long-lived particle sensitivity studies}}
}


\author[c]{Sofie Nordahl Erner}
\author[d]{Anna Mullin}
\author[b]{\texorpdfstring{Th\'eo Reymermier\corref{author}}{Th\'eo Reymermier}}
\author[a]{Paul Swallow}
\author[a]{Oleg Brandt}
\author[]{for~the~ANUBIS~Collaboration}

\cortext[author] {Corresponding author.\texorpdfstring{\\\textit{E-mail address:}}{E-mail address:} theo.reymermier@cern.ch}

\address[a]{Cavendish Laboratory, University of Cambridge, Cambridge, United Kingdom}
\address[b]{Universit\'e de Lyon, Universit\'e Claude Bernard Lyon 1, CNRS/IN2P3, Institut de Physique des 2 Infinis (IP2I), Lyon, France}
\address[c]{Formerly at: IPPP, Dept of Physics, Durham University, Durham, United Kingdom}
\address[d]{Formerly at: Cavendish Laboratory, University of Cambridge, Cambridge, United Kingdom}

\begin{abstract}
The proposed ANUBIS detector has been designed to search for Long-Lived Particle (LLP) signatures, which have become a focus for a variety of experiments within recent years. Due to the variety of possible LLP models and the need for directly comparable sensitivity results highlighting the potential coverage of ANUBIS, the SET-ANUBIS (Simulation, accEptance and sensiTivity studies framework for ANUBIS) framework has been developed.

SET-ANUBIS is a flexible open-source Python framework that can perform full sensitivity studies of LLP signatures for the ANUBIS detector, allowing others to directly produce ANUBIS-like limits or even implement other geometries and detectors. It starts from a Universal FeynRules Output model or user-supplied rates; it exposes model parameters and particle content for user modification, and evaluates decay widths, branching ratios and lifetimes. \marty can also be used to compute matrix elements, decay widths, branching ratios, and cross-sections. Then the framework prepares or runs event generation through \pythia{8} or \madgraph, ingests HepMC event records, models the geometry of the ATLAS cavern and ANUBIS tracking stations, and applies a configurable sequence of geometric, kinematic and isolation requirements to select a set of surviving LLP candidates for evaluation. The implementation follows a ports-and-adapters architecture so that the core logic remains separate from external generators, persistent storage and visualisation. A scan-aware SQLite catalogue and content-addressed store preserve cards, metadata and compact selection-ready event bundles while avoiding duplicate artifacts. A pre-release version of the SET-ANUBIS framework has already been used to successfully derive the sensitivity of ANUBIS to three LLP benchmark models involving a Higgs portal and a Heavy Neutral Lepton model.
 
   \begin{center}
       \today 
   \end{center}
   \vspace{-0.5cm}
\end{abstract}

\begin{keyword}
Long-lived particles \sep ANUBIS \sep Pythia8 \sep MadGraph \sep MARTY \sep sensitivity studies \sep hexagonal architecture.
\end{keyword}

\end{frontmatter}




\onecolumn
\clearpage
\noindent
{\bf PROGRAM SUMMARY}

\noindent
{\em Program Title:} \setanubis \\
{\em CPC Library link to program files:} To be added by the Technical Editor \\
{\em Developer's repository link:} \setanubisrepo \\
{\em Documentation:} \setanubisdocs \\
{\em Release identifier:} \setanubistag \\
{\em Archived release:} \setanubisdoi \\
{\em Package distribution:} \setanubispypi via PyPI after TestPyPI checksum and installation verification \\
{\em Licensing provisions:} GNU General Public License v3 or later (\texttt{GPL-3.0-or-later}) \\
{\em Programming language:} Python ($\geq 3.10$); C++ for the optional native \pythia binding; Bash/Docker for external generator workflows \\
{\em Supplementary material:} Reproducibility scenarios R1--R5, versioned inputs and expected outputs, a compact HepMC2 event sample, JSON and standalone HTML selection traces, examples and the user manual. \\
{\em Nature of problem:}
Sensitivity studies for Long-Lived Particles (LLPs) require several software domains to agree on model parameters, decay information, generated events, detector coordinates and event-selection definitions. These steps are commonly split across independent tools and file formats, making comprehensive scans of model parameters difficult to carry out, audit, and reproduce. The problem addressed by \setanubis is to connect these operations through a consistent interface while retaining the original parameter cards, model values, event representations, selection configuration and provenance required to reconstruct an \anubis acceptance calculation. \\
{\em Solution method:}
\setanubis organises the workflow as bounded domains with stable ports and replaceable adapters. Universal FeynRules Output models and user calculators provide particle and decay information; deterministic builders prepare \pythia command files and \madgraph cards; generated HepMC records are converted to compact pandas dataframes; geometry predicates describe the \atlas cavern geometry and \anubis tracking layers; and a named cutflow evaluates LLP decay, geometric acceptance, \atlas-veto, tracking, missing transverse momentum and isolation requirements. A content-addressed store and SQLite catalogue deduplicate artifacts and preserve campaign metadata. Reproducibility scenarios exercise each layer independently and compare generated summaries with versioned references. \\
{\em Additional comments including restrictions and unusual features:}
The core model, card-generation and selection APIs are pure Python. Running \pythia itself requires the optional compiled \texttt{pythia\_sim} binding, while \madgraph and \marty require their respective external installations or containers. R3 and R4 intentionally validate deterministic command/card construction without invoking those external engines. The two Dash applications are optional inspection tools and do not modify production cutflows. Linux is the primary supported platform; documented WSL and container workflows are provided for other host systems.
\twocolumn
\clearpage

\section{Introduction}

\setanubis (Simulation, accEptance and sensiTivity studies framework for \anubis) is a framework built to perform sensitivity projections for the \anubis detector~\cite{Bauer:2019vqk,ANUBIS:2025sgg} across multiple Beyond the Standard Model (BSM) scenarios with Long-Lived Particles (LLPs). The framework connects to established HEP tools such as \pythia{8}~\cite{Sjostrand:2014zea} and \madgraph~\cite{Alwall:2014hca} to perform event generation, or \marty~\cite{Uhlrich:2020aaj} to compute scattering matrix elements, decay widths, branching ratios (BRs) and cross sections. Multiple abstraction layers are available to switch between different tools or calculation methods, as well as performing a scan over a given set of parameters in a BSM theory.

The pipeline is separated into parts that are connected through a hexagonal (`ports-and-adapters') architecture. In this structure, there is a domain, which defines a narrow, explicit API, and there are adapters, which are external scripts and I/O for the user. Ports are the interfaces between the adapters and the domain.  This arrangement keeps coupling low and lets a generator, a database, or a geometry backend be swapped without changing domain code (and vice versa). External tools can be installed within the pipeline, such as \madgraph, but it is also possible to run \madversion{3.5.8} inside a Docker container through an interface in the pipeline to avoid system-level conflicts.

This paper gives an overview of the physics context (Section~\ref{sec:Physics}) and the methodology of the program (Section~\ref{sec:Methodology}). It focuses on the architecture and main parts of the framework (Section~\ref{sec:Architecture}), and on how the parts interact in an end-to-end analysis, \ie covering each step from start to finish. This includes model and parameter handling; decay width and branching ratio evaluation; generation and ingestion of events; modelling the \atlas cavern and \anubis station geometry; and the selection pipeline that decides whether an LLP decay would yield detectable activity in \anubis. Section~\ref{sec:ValidationAndExamples} details examples based on the Standard Model and a Heavy Neutral Lepton (HNL) benchmark that demonstrate the public interfaces, while automated tests and five reference scenarios validate the release. Executable examples and Google-style docstrings are also distributed with the package. Reproducibility is supported by versioned inputs and expected outputs; a Content-Addressed Store (CAS), \ie storing the data as a unique cryptographic hash of the content; and an SQL catalogue that deduplicates generated artifacts by stable hashes, where an artifact here refers to the outputs of the framework (Section~\ref{sec:Reproducibility}). The performance of the framework, as well as its limitations, is discussed in Sections~\ref{sec:Performance} and~{\ref{sec:Limitations}} respectively, before concluding in Section~\ref{sec:conclusion}.

\section{Physics summary}
\label{sec:Physics}
Long-lived particles, \ie particles with $\tau>10$~ps, are a generic feature in many theoretical models of new physics beyond the Standard Model (SM) of particle physics~\cite{Alimena:2019zri}. The main focus of \anubis is on LLPs with decay lengths of $\mathcal{O}(1)$~m and above.

The \atlas~\cite{ATLAS:2018tup} and CMS~\cite{CMS:2014hka,CMS:2021sch} experiments are performing searches for LLPs. However, they are fundamentally limited by their physical size and location around an interaction point (IP) of the LHC, so a neutral LLP could pass completely undetected through the detector before decaying outside. Their trigger and reconstruction algorithms are moreover optimised for objects originating close to the IP. Therefore, dedicated displaced LLP search experiments have been proposed to cover areas of the proper lifetime parameter space not accessible to the existing detectors~\cite{Bauer:2019vqk,Gligorov:2017nwh,Curtin:2018mvb,SHiP:2021nfo}.

\begin{figure}[ht]
    \centering
    \begin{subfigure}{0.49\linewidth}
    \centering
    \includegraphics[width=\linewidth]{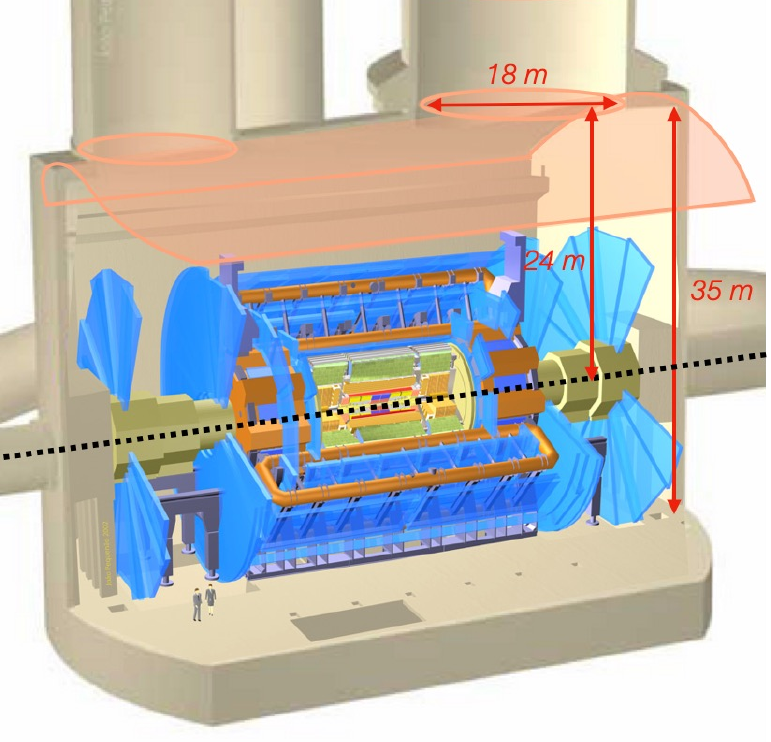}
    \caption{}
    \label{fig:SketchATLAS_undergroundCavern}
    \end{subfigure}
    \begin{subfigure}{0.49\linewidth}\includegraphics[width=\linewidth]{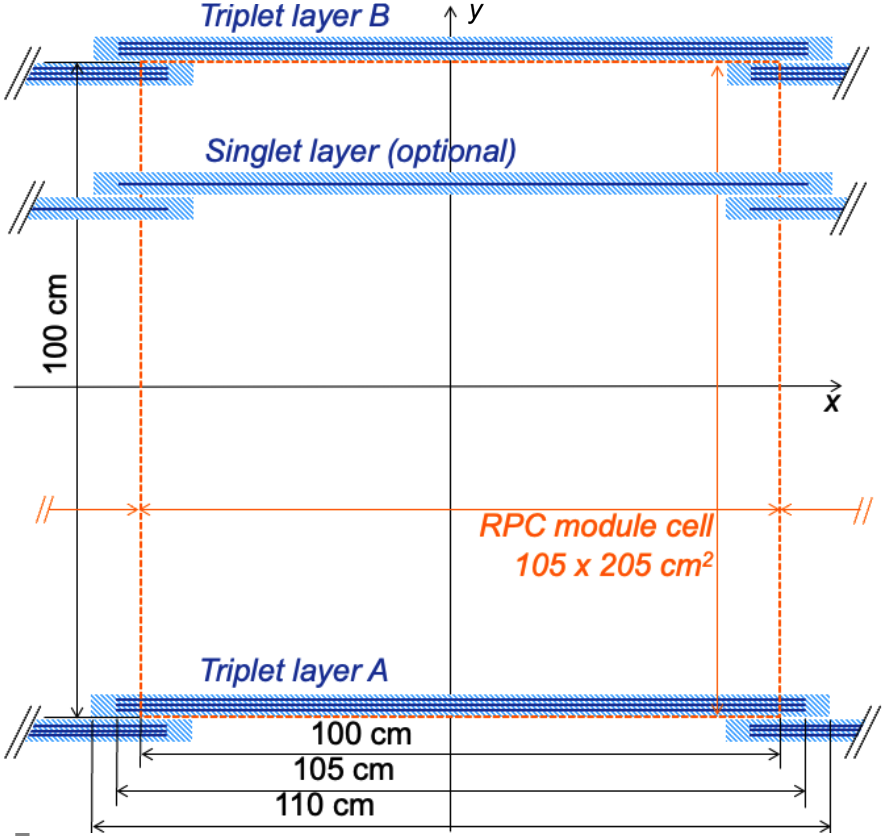}
    \caption{}
    \label{fig:RPCLayout}
    \end{subfigure}
    \caption{(a) The layout of the underground cavern at LHC Point 1, UX15, featuring the \atlas experiment and the PX14 and PX16 service shafts above it. The proposed configuration of the \anubis detector is shown in orange. This includes two circular stations at the bottom of the two access shafts. (b) Layout of the RPC tracking layers for the \anubis detector, with two layers of triplets separated by $\sim1~\metre$ and an optional singlet layer installed between them.}
    \label{fig:ANUBISDetector}
\end{figure}

\anubis is a proposed LLP detector that would be located on the ceiling of the UX15 \atlas Experimental Cavern at CERN, instrumented by a set of Resistive Plate Chambers (RPCs) as the tracking technology. It is expected to have a proper lifetime, $c\tau$, reach ranging from $\order{1}~\metre$ to $\mathcal{O}(10^5)~\metre$, and by working symbiotically with \atlas extend the coverage from $\mathcal{O}(10^5)~\metre$ down to prompt decays at the IP~\cite{Bauer:2019vqk,ANUBIS:2025sgg}. Figure~\ref{fig:SketchATLAS_undergroundCavern} shows the geometry of the \anubis detector within the \atlas cavern. It is composed of two main tracking layers separated by $1~\metre$, as shown for a single module in
Figure~\ref{fig:RPCLayout}. These modules are tiled such that they follow the curve of the ceiling. These layers contain three individual RPC detectors, and there is an optional singlet layer that is being considered that would be installed between them to improve vertex reconstruction and disambiguation of tracks. The RPCs are only sensitive to charged particles,
so when considering the sensitivity of \anubis, an LLP candidate is required to have at least two charged tracks that intersect both of the main tracking layers with a vertex that lies in the air volume between \atlas and \anubis.\\

A pre-release version of the \setanubis framework has already been used to successfully derive the sensitivity of the ANUBIS detector to several LLP models~\cite{ANUBIS:2025sgg,Reymermier:2026HNL,PBC:2025sny}. In particular, it was used to examine a set of benchmark cases (BCs) proposed by the Physics Beyond Colliders collaboration to create direct comparisons between different LLP experiments~\cite{Beacham:2019nyx,Alemany:2019vsk}.
The development of the framework and these studies occurred concurrently, informing the necessary features and identifying any issues with \setanubis before release. Therefore, there may be some difference between the generator versions used previously and reported here. 

\section{Program overview and methodology}
\label{sec:ProgramOverview}
\label{sec:Methodology}
\setanubis has been designed to perform sensitivity studies for the \anubis detector for an arbitrary BSM model provided by the user with minimal additional bespoke code required. By providing a flexible, modular framework that is common for all the models considered, it enables better reproducibility of results and direct comparison since all models are processed with the same tools.

To achieve this, \setanubis is organised as a set of bounded contexts with hexagonal boundaries, following a hexagonal architecture. The aim of using this architecture for the framework is to keep domain APIs small and orthogonal with low coupling to other parts, isolate I/O and external toolchains in adapters, and provide a complete end-to-end pipeline, from defining the model parameters using the Universal FeynRules Output (UFO)\footnote{The framework is also compatible with the updated Universal Feynman Output (UFO 2.0) format~\cite{Darme:2023jdn}.} format~\cite{Degrande:2011ua}, and selecting events that are expected to be observed in \anubis-like geometries. Reproducibility is addressed with a CAS for heavy artifacts and an SQLite database in WAL (Write-Ahead Logging) mode.


There are three main aspects to the framework: Simulation of signal samples; geometric acceptance and event selection; and determination of the sensitivity limits.

\subsection{Simulation of signal samples}
\label{sec:signalSimulation}
For the simulation, several Monte Carlo (MC) generators are supported, so that the generator best suited to a given model can be chosen.
The only requirement is that the acceptance and event selection in the framework assumes the produced samples are in the HepMC file format~\cite{Buckley:2019xhk}, which is a common file format output used by many MC generators to ensure interoperability. Currently, the baseline version of \setanubis includes methods for using \madgraph~\cite{Alwall:2014hca}, specifically \madversion{3.5.8}, and \pythia{8}~\cite{Bierlich:2022pfr}. The architectures of these parts of the framework are described in Section~\ref{sec:Madgraph} and~\ref{sec:PythiaInterface} respectively.

For \madgraph the implementation of LLP models uses the FeynRules~\cite{Alloul:2013bka} package and Universal FeynRules Output model files. The UFO files contain all relevant information on the dynamics for an LLP model by implementing the associated Lagrangians, allowing for the calculation of the production and decay branching ratios of the modes defined in that model for a set of given model parameters. These UFO files are a common tool, with many typical benchmark physics models already having readily available files used by many different experiments. Through the use of \madgraph and \madspin, signal samples containing both the production and decay of LLPs can be produced at the partonic level. Typically, \pythia is also used to hadronise the final states to accurately simulate the visible decay products.

To simulate signal events with \madgraph, a `jobscript' is defined that contains the commands to be called within \madgraph, including the definition of the LLP production and a set of `cards' (text files that are input to the generator). These contain: a `param' card, outlining the SM and BSM parameters; a `run' card, that gives the configuration of the run \eg the beam conditions, the PDF set, the number of events to produce, and the random number seed; a \madspin card, that defines subsequent decays and can handle more than the default 2-body cases \madgraph can; and finally a \pythia card, that handles the hadronisation of the final states. The framework allows for a template jobscript to be created for particular production modes and the \madspin cards that define the decay modes, run cards with LHC conditions and a unique random seed for each run, and then create a complete \madgraph jobscript with these combined, which can also specify an output directory. Additionally, in the jobscript \madgraph supports scans of model parameters, \eg LLP mass, LLP coupling \etc, which can also be specified when the framework produces jobscripts.

These jobscripts can then be run via the framework running \madgraph as an external tool or in a Docker environment, or by the user submitting them to a batch system, such as HTCondor~\cite{bib:condor}, where \madgraph is installed. The simulated signal events are then output to the specified directory with the format:
\begin{lstlisting}[language=bash]
<output directory>/
  Events/
    run_01/
      event.hepmc
      run_01_tag_1_banner.txt
    run_02/
      ...
  scan_run_[01-N].txt
\end{lstlisting}
where `scan\_run\_[01-N].txt' is a summary file that is produced when parameter scans are set up in the jobscript. It contains a table with the parameters that were scanned, the associated decay width and value of the cross-section\footnote{When using \madspin the produced $\sigma$ value, labelled as `cross' in the table and the summary text file, in fact represents $\sigma\times\BF$, where $\BF$ is the branching ratio of the decay mode simulated.}, $\sigma$. This provides a useful starting point for a database that can associate a particular sample with the model parameters used to produce it. The database layer of the framework enables the parsing of the \madgraph file structure to do this, see Section~\ref{sec:Database}.\newline

Instead of allowing \madgraph to assign the branching ratio of the generated process automatically through its own calculations, it can be set manually. The MARTY~\cite{Uhlrich:2020aaj} framework calculates all the scattering matrices, matrix elements, decay widths and cross-sections using Feynman diagrams such that the derived branching ratios should align with the theoretical formulations of the model without simplifying assumptions, which \madgraph occasionally has to apply. \setanubis provides an interface with MARTY to provide the user with the option to evaluate the branching ratios explicitly, and potentially cross-check the branching ratios in \madgraph to ensure that they are being handled correctly.\newline

To simulate signal events with \pythia, the generator can be controlled via a command (\texttt{.cmnd}) file that is comparable to the jobscript for \madgraph in that it defines the configuration of the run, including beam conditions, and the definition of new decay modes with particular branching ratios and cross-sections. Here, the branching ratios for the new LLP model are not calculated explicitly within \pythia, like with the UFO files in \madgraph. Instead, the branching ratios need to be provided in the .cmnd file before the run. The framework allows for these to be provided in several ways (see Section~\ref{sec:BranchingRatio}): via a table of calculated values, to provide flexibility; through user-created Pythonic functions; through \marty; or even via \madgraph with a given UFO file. Once the framework produces the .cmnd file for the signal sample with a particular set of model parameters it can then be run with \pythia. The output HepMC file is then put into a similar file structure as the one provided by \madgraph, including the creation of a scan text file to allow for ingestion into the framework's database with the same information as \madgraph produced samples.

Additional MC generators can be implemented into the framework upon request from users to ensure that the produced samples can then be added to the database properly. The defined database structure provides robust documentation when there are $\mathcal{O}(10^3)$ or more samples produced with different parameters, and to provide a common and defined structure for use by other aspects of the framework.

\subsection{Geometric acceptance and event selection}
Once signal events are generated, the next stage is to determine the subset of those events that the \anubis detector could potentially observe. This is done by imposing a set of selections, known as a `cutflow', which include geometric and kinematic considerations using the Selection part of the framework, described in Section~\ref{sec:selection}. To maintain the modular nature of the framework and make it possible for other non-\anubis geometries to be implemented, these selections are imposed via a set of functions for each step and the full cutflow is defined by a collection of these functions. The HepMC file with a signal sample is converted by the framework into a pandas dataframe to allow for more efficient slicing of the events. This dataframe is then processed by using the event information to create useful variables such as the pseudorapidity, $\eta$, and $\phi$ angle in the \atlas coordinate system\footnote{%
\atlas adopts a right-handed coordinate system, where its origin is at the nominal interaction point in the centre of the detector and the $z$-axis points along the beam pipe. The $x$-axis points from the IP to the centre of the LHC ring, and the $y$-axis points upwards.
Polar coordinates ($r$, $\phi$) are used in the transverse plane, $\phi$ being the azimuthal angle around the $z$-axis. 
The pseudorapidity is defined in terms of the polar angle $\theta$ as $\eta \equiv-\ln \tan(\theta/2)$ and is equal to the rapidity $y = \frac12 \ln \left(\frac{E+p_z}{E-p_z}\right)$
in the relativistic limit.
Angular distance is measured in units of $\Delta R \equiv \sqrt{\Delta\eta^2+\Delta\phi^2}$.
}, and then separating it into a set of sub-dataframes for particular particle objects: charged final state particles, neutral final state particles, LLP particles and the LLP decay products.

The definition of the charged and neutral final state particles can be controlled by a given configuration dictionary. In the nominal configuration these are considered as prompt final-state particles in the signal simulation sample that are a small radial distance from the primary IP, \ie $d_{IP}<10\,\text{mm}$, and are not produced by the LLP candidate. These have a non-zero or zero electric charge for the neutral and charged final states respectively, and the charged state also requires an additional selection of $\pt>5$~GeV. Then `jets' are created from jet candidates composed of all prompt final-state particles not produced by the LLP and are not neutrinos, which are then passed to the anti-$k_t$ algorithm~\cite{Cacciari:2008gp} with $R=0.4$. Therefore, a charged-particle object may also be a jet constituent. These definitions are summarised in Table~\ref{tab:PartDefinition}. The LLPs are identified using their PDG ID number~\cite{ParticleDataGroup:2024cfk}, and their decay products are found by recursively searching the decay products for each LLP. These processed dataframes can then be cached before the selection is applied.

\begin{table}[ht]
\caption{Summary of the definition of objects considered in the selections. 
The jets are reconstructed using the anti-k$_t$~\cite{Cacciari:2008gp} jet algorithm with $R=0.4$ from prompt final-state particles that pass the selections outlined below.
}
\label{tab:PartDefinition}
\resizebox{\columnwidth}{!}{
\begin{tabular}{ll}
\hline

\textbf{Objects} & \textbf{Selections} 
\\ 
\hline
\hline
\multirow{2}{*}{Prompt charged particles}\quad\hbox{}     & $q\neq0$, $d_\text{IP}<10~\mm$, \\
 & $|\vec p|>0.1~\GeV$, $\pt>5~\GeV$ \vspace{0.15cm}\\
\hline
\multirow{3}{*}{Prompt particles}     & $d_\text{IP}<10~\mm$, \\
& $|\vec p|>0.1~\GeV$, \\
& ID$_\text{PDG}\notin[12,14,16,18]$ \vspace{0.15cm} \\ \hline
Jets                  & $\pt>15~\GeV$ \\ \hline
\end{tabular}
}
\end{table}

In the nominal cutflow the geometric acceptance is considered first. This is implemented through a set of boolean predicates, \ie whether the decay position of the LLP lies within a certain region or not; these are referred to as `is\_in' functions. For this, a simplified \texttt{ATLASCavern} class was created that contains information on the \atlas UX15 cavern dimensions, including the radius of curvature of its vaulted ceiling, location of the PX14 and PX16 service shafts, as well as the location of the IP, which is offset from the centre of curvature of the ceiling and the centre of the cavern. Additionally, the dimensions of \atlas and a way to define a set of partial cylindrical shells to replicate \anubis' tracking layers attached to the ceiling are also defined. The number of \anubis tracking layers, their vertical separation, the number of RPCs per layer and the RPCs' efficiency can all be defined by the user\footnote{Though these are referred to as RPCs in the context of \anubis, these can be treated generically as individual tracking planes within a tracking layer.}. In the nominal case, the geometry is defined by two tracking layers separated by 1~\metre with the top layer 0.2~\metre below the cavern ceiling following its curvature, each layer contains a single RPC, and the RPCs are set to be 100\% efficient. An LLP event is considered to be potentially detectable by \anubis if its decay vertex lies in the air-filled volume between \atlas and \anubis. This is imposed in several stages: first, the position of LLP decay vertex is checked if it lies within the \atlas cavern, and then whether it falls into the region between 9.5~\metre from the \atlas IP and 0.3~\metre below the bottom tracking layer of \anubis in the radial direction while simultaneously fulfilling $|z|<22$~\metre. After this the LLP decay products are considered by projecting the charged tracks of the LLP decay. If they leave at least two charged tracks within \anubis' tracking layers and each intersects both of the tracking layers, then the LLP candidate passes the acceptance selection.
In the next step, a set of kinematic selections is imposed; these are designed to replicate the background rejection strategy that would be applied to the \anubis data set by exploiting the association with \atlas. 
For example, a key signature of an unrecorded neutral LLP candidate in the \atlas detector is the presence of missing transverse momentum, \met or `MET', in an event. 
The main backgrounds that are expected for \anubis are energetic long-lived neutral kaons $K_L$ and neutrons $n$ that escape the \atlas detector and interact hadronically with the material inside the \atlas cavern, typically nuclei of air molecules~\cite{ANUBIS:2025sgg}. 
Such $K_L$ and $n$ can be produced either promptly at the IP, typically in association with other hadrons, or from energetic hadronic jets that punch through the ATLAS calorimeters. 
Both processes introduce small amounts of \met that typically correspond to a few \GeV. 
Therefore, a selection of ${\met>30\,\text{GeV}}$ is applied to eliminate such backgrounds.
Finally, there is a set of isolation requirements designed to remove the remaining background using the fact that for both aforementioned processes the $K_L$ and $n$ are produced in association with hadronic activity that is closeby in $\Delta R$.
Isolation requirements are imposed by requiring the LLP trajectories to have an angular separation of $\Delta R_\text{LLP,jet}>0.5$ from any jets with $\pt>15~\GeV$ and of $\Delta R_\text{LLP,trk}>0.5$ from any charged particles with $\pt>5~\GeV$.
After applying the above selections, known as the \atlas veto, the expected background level is strongly reduced~\cite{ANUBIS:2025sgg}.

\subsection{Determination of the sensitivity limits}
The final analysis stage supported by \setanubis is to obtain the projected sensitivity limits. Generally, to consider if a model would be observable at \anubis it is required that the number of observed LLP events, $N_{LLP}$, significantly exceeds the background estimation at a 95\% confidence level (CL), or in an assumed background-free scenario $N_{LLP}\geq4$. The conservative background estimation for \anubis, as considered in Ref.~\cite{ANUBIS:2025sgg}, corresponds to $N_{LLP}\geq36$ for a 95\% exclusion sensitivity with the $CL_\text{s}$ method~\cite{Read:2002hq}. In \setanubis the required observation threshold on $N_{LLP}$ can be assigned by the user, allowing for additional background scenarios to be considered.

The value of $N_{LLP}$ can be determined by
\begin{align}
    N_{LLP} = \mathcal{L}\,\sigma_{LLP}\,\BF_{LLP}\,\mathcal{A}\,\varepsilon_{sig},
    \label{eq:NLLP}
\end{align}
where $\mathcal{L}=3\,\text{ab}^{-1}$ is the integrated luminosity that is expected to be delivered by the High-Luminosity LHC (HL-LHC) operating at $\sqrt{s}=14~\TeV$; $\sigma_{LLP}$ is the production cross-section of the LLP; $\BF_{LLP}$ is the branching ratio of the LLP into the considered final state(s); $\varepsilon_{sig}$ is the efficiency with which \anubis could observe a signal event, and is usually taken as 50\%; and $\mathcal{A}$ is the kinematic and geometric acceptance of \anubis which is defined as the fraction of LLP events that pass the cutflow outlined previously and decay within the active volume of ANUBIS, compared to the total number of generated events. From Equation~\eqref{eq:NLLP}, there are several variables which will depend on the parameters of the LLP model used in the simulated samples; from this the sensitivity limits can be cast into different parameters of interest. Commonly, this is done by setting $N_{LLP}$ to the observation thresholds, \eg 4, and then calculating the associated value of $\sigma_{LLP}$, $\BF_{LLP}$, or $\sigma_{LLP}\,\BF_{LLP}$ for each of the relevant scan parameters, \eg $m_{LLP}$. Alternatively, $N_{LLP}$ can be determined directly by taking the simulated values for each term in Equation~\eqref{eq:NLLP} for each combination of $N$ model parameters to create an $N$-dimensional histogram with each bin set to the associated $N_{LLP}$ value, and the sensitivity limits defined by the $N_{LLP}$ contour for the observation threshold.

This determination can get more complicated with the LLP model complexity, but the modular nature of \setanubis ensures that Equation~\eqref{eq:NLLP} can be adapted for each model. However, the basis of the sensitivity limit should remain similar to allow for easier comparison of limits from different LLP models.

\section{Software architecture}
\label{sec:Architecture}
Figure~\ref{fig:global-architecture} shows the global architecture of the framework, where each part follows the hexagonal architecture, except the \textit{UFOInterface} and \textit{Common} parts. Each domain (core) implements pure logic and exposes a small interface, without any external dependencies. Adapters implement concrete integrations, for example parsing a UFO model, calling event generators, reading and writing event artifacts or plotting. This layout allows \setanubis to remain stable as backends evolve, because only adapters need to change. A summary of the adapters, domain and ports for each part of the framework is given in Table~\ref{tab:Architecture}.

\begin{figure*}[ht]
  \centering
  \includegraphics[width=0.77\linewidth]{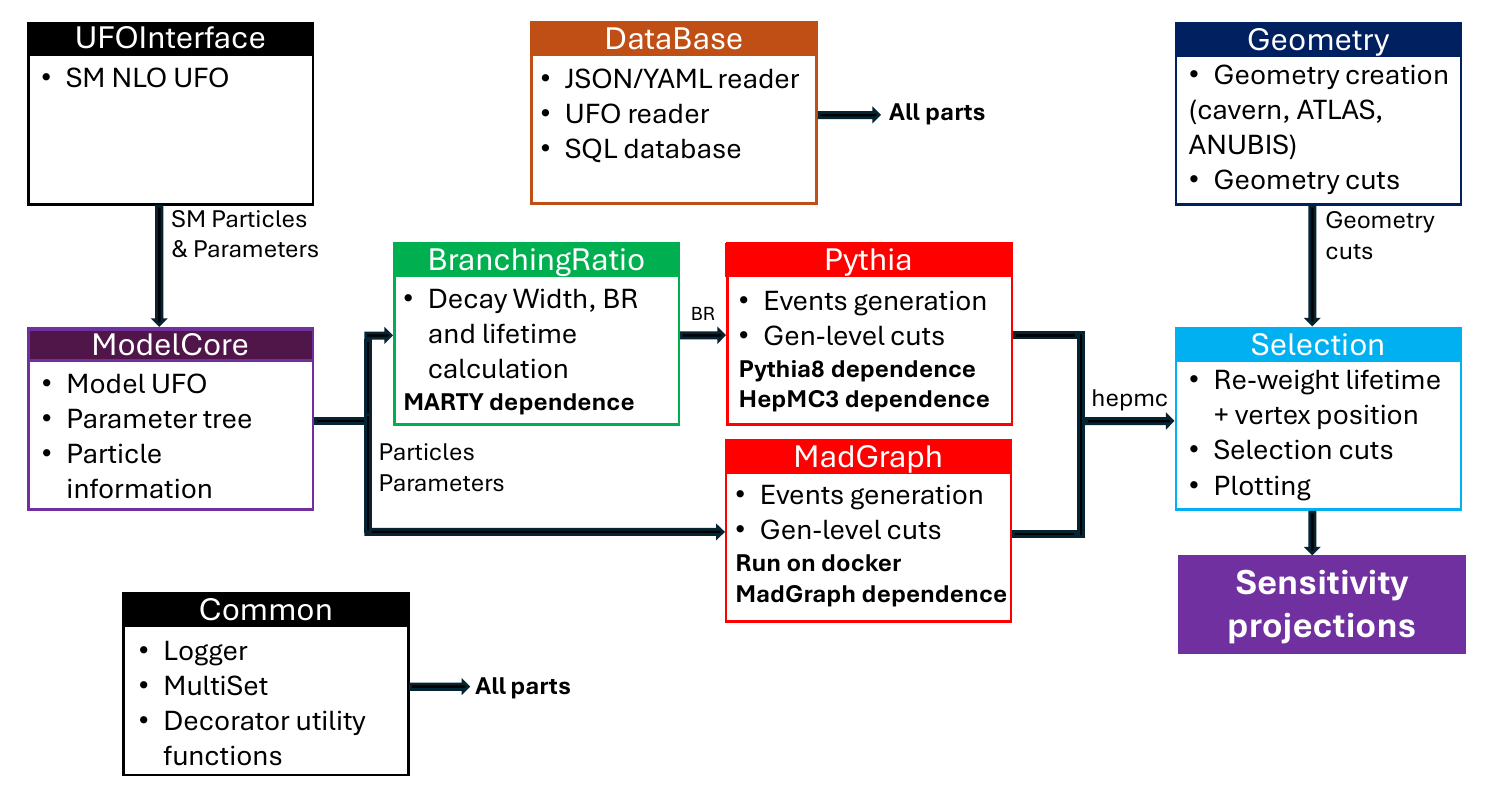}%
  \caption{Global architecture of \setanubis. Domain kernels are in the centre and adapters connect to external tools such as \pythia{8}, \madgraph, \marty, the event database and geometry representation.}
  \label{fig:global-architecture}
\end{figure*}

\begin{table*}[htb]
\caption{Summary of the adapters, domain and ports for each of the different layers of the \setanubis architecture.}
\label{tab:Architecture}
\resizebox{\linewidth}{!}{%
\begin{tabular}{llll}
\hline
\textbf{Layer} &
  \textbf{Adapters} &
  \textbf{Domain} &
  \textbf{Ports} \\ \hline
DataBase &
  \begin{tabular}[c]{@{}l@{}}\textbf{Input}: CardGetter,  DecayProvider,  \\ FileCopyBuilder,  JSONExtractor,\\ ParamCardGeneratorAdapter, \\ ParamProvider, ParticleProvider, \\ UFOInterface, UFOParser, YAMLParser.\end{tabular} &
  \begin{tabular}[c]{@{}l@{}}EventDataBaseManager, \\UFOManager,  CardGenerator.\end{tabular} &
  \begin{tabular}[c]{@{}l@{}}\textbf{Input}: MadGraph (ICardGetter), \\ ModelCore (IUFOParamInterface).\end{tabular} \vspace{0.1cm} \\ \hline
ModelCore &
  \begin{tabular}[c]{@{}l@{}}\textbf{Input/output}: SetAnubis interface, \\ QCD adapter, UFO getter.\end{tabular} &
  \begin{tabular}[c]{@{}l@{}}SetAnubisManager, QCDRunner, \\ Parameter tree evaluation, particles catalogue, \\ QCD running and $\alpha_s$ helper.\end{tabular} &
  \begin{tabular}[c]{@{}l@{}}\textbf{Output}: DataBase (ExpressionTree).\\ \textbf{Input}: Every other part (SetAnubisInterface).\end{tabular} \vspace{0.1cm} \\\hline 
BranchingRatio &
  \begin{tabular}[c]{@{}l@{}}\textbf{Input/output}: DecayInterface, \\ UFO/Python/File/\madgraph/\marty.\end{tabular} &
  \begin{tabular}[c]{@{}l@{}}BranchingRatioManager, \\ IDecayCalculation,  IDecayChecker,  \\ Calculations of decay widths, BRs and lifetimes, \\ Registration of the decay channels, \\ Validity checks (charge conservation).\end{tabular} &
  \begin{tabular}[c]{@{}l@{}}\textbf{Output}: DataBase (DecayProvider).\\ \textbf{Input}: Pythia (DecayInterface).\end{tabular} \vspace{0.1cm} \\ \hline 
MadGraph (interface) &
  \begin{tabular}[c]{@{}l@{}}\textbf{Input/output}: JobscriptBuilder, \\ RunCardBuilder, MadSpinCard Adapter, \\ ParamCardInitializer.\end{tabular} &
  \begin{tabular}[c]{@{}l@{}}MadGraphCommandCard, RunCard editor,  \\ MadSpin builder.\end{tabular} &
  \begin{tabular}[c]{@{}l@{}}\textbf{Output}: DataBase (CardAdapter).\\ \textbf{Input}: UI (Builders and Adapters).\end{tabular} \vspace{0.1cm} \\\hline 
Pythia (interface) &
  \begin{tabular}[c]{@{}l@{}}\textbf{Input}: PythiaCMNDInterface, \\ CMNDScanManager, PythiaRunInterface\end{tabular} &
  \begin{tabular}[c]{@{}l@{}}CMNDGenerationManager, \\ CMNDSection Particle/Decay formats.\\ \textbf{External Dependencies}: \pythia{8},\\ through a pybind11 python binding.\end{tabular} &
  \begin{tabular}[c]{@{}l@{}}\textbf{Output}: DataBase, BranchingRatio.\\ \textbf{Input}: User.\end{tabular} \vspace{0.1cm} \\\hline 
Geometry &
  \begin{tabular}[c]{@{}l@{}}\textbf{Input/output}: CavernGeometryBuilder, \\ GeometrySelectionAdapter\end{tabular} &
  \begin{tabular}[c]{@{}l@{}}ATLASCavern, \\ Definition of the \anubis RPC layers, \\ RPC detection efficiency etc.\end{tabular} &
  \begin{tabular}[c]{@{}l@{}}\textbf{Output}: DataBase (YAML reader).\\ \textbf{Input}: Selection (`is\_in\_' adapters).\end{tabular} \vspace{0.1cm} \\ \hline
Selection &
  \begin{tabular}[c]{@{}l@{}}\textbf{Input/output}: SelectionPipeline, \\ EventsDBHepmcSelectorAdapter \etc\end{tabular} &
  \begin{tabular}[c]{@{}l@{}}SelectionEngine, SelectionConfig,  RunConfig, \\ IsolationComputer, JetDFBuilder, \\ LLPAnalyzer\end{tabular} &
  \begin{tabular}[c]{@{}l@{}}\textbf{Output}: Geometry (geometry cuts).\\ \textbf{Input}: UI (Pipeline).\end{tabular} \\ \hline
\end{tabular}%
}
\end{table*}

\subsection{Common \& UFOInterface}
\label{sec:CommonAndUFOInterface}
These parts do not follow the hexagonal architecture, as their only use is to define structures that can be used in every other part of the framework \eg \textit{MultiSet}, special decorators, and the logger. This is shown in Figure~\ref{fig:commonarchi}, alongside the structure of the whole framework. 

\textit{Common} hosts cross-cutting utilities such as the logger and a minimal set of decorators for tracing, timing, validation, caching and retries, along with a small \texttt{MultiSet} data structure. \texttt{MultiSet} is used such that an array of particle IDs is treated independently of its ordering, \eg [13,-13] is identical to [-13,13].

\textit{UFOInterface} provides a baseline Standard Model UFO used by other parts with all SM particles and parameters included by default. SM Parameters (when they are not modified in the BSM convention) will be automatically calculated in the pipeline and used to simplify the sympy~\cite{bib:meurer2017sympy} expression of decay width, reducing the evaluation cost.

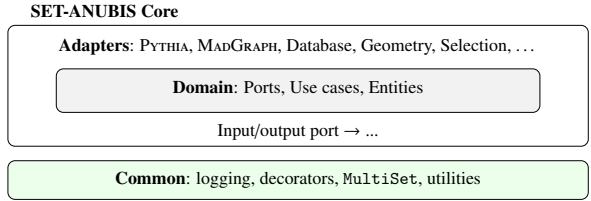
\begin{figure}[htb]
\begin{center}
\resizebox{\linewidth}{!}{
\begin{tikzpicture}[scale=0.95]
  \node at (-4,3.3) {\textbf{\setanubis Core}}; 
  \draw[rounded corners] (-6,0.5) rectangle (6,3);
  \node at (0,2.55) {\textbf{Adapters}: \pythia, \madgraph, Database, Geometry, Selection, \ldots};
  \draw[rounded corners, fill=gray!10] (-5,1.2) rectangle (5,2.1);
  \node at (0,1.7) {\textbf{Domain}: Ports, Use cases, Entities};
  \node at (0,0.8) {Input/output port $\rightarrow$ ...};
  \draw[rounded corners, fill=green!8] (-6,-0.6) rectangle (6,0.2);
  \node at (0, -0.2) {\textbf{Common}: logging, decorators, \texttt{MultiSet}, utilities};
\end{tikzpicture}
}
\end{center}
\caption{General structure of the \setanubis framework, showing the relation between the adapters, domain, and ports of the core, as well as the place of the Common layer outside the hexagonal boundaries of the core.}
\label{fig:commonarchi}
\end{figure}

\subsection{DataBase}
\label{sec:Database}
The \textit{DataBase} part provides parsing services and event storage. It parses UFO models and builds an expression tree for parameters and decays defined in the model, which can then be evaluated. It can also generate and retrieve templates of the \madgraph's job script and text cards, like the \pythia card for the showering, \madspin card for the LLP's decay and the `param' card plus the run card for the execution of \madgraph.

The database layer can also generate a template of a \texttt{.cmnd} file for \pythia, and ingests event runs and scan tables. The event store combines a SQLite catalogue with a CAS area keyed by SHA-256 hashes. SQL tables track models, events, artifacts and binary large objects (blobs), while forward-compatible migrations add scan-related columns on demand, \eg for finding simulation samples for a particular model with a specific set of input parameters.
In the current storage model, the importer reads the HepMC stream, constructs a tabular event representation, \ie a pandas dataframe, with \texttt{HepmcFrameBuilder}, applies the \texttt{LLPAnalyzer}, and writes a compressed \texttt{dict[str, pandas.DataFrame]} bundle. The dataframe contains a set of columns derived from the information in the HepMC file, which can be used for selection, these are shown in Table~\ref{tab:pandasCol}. 
\begin{table*}[htb]
\caption{Summary of the columns of the pandas dataframe within the event bundle used for the \texttt{LLPs}, \texttt{LLPchildren}, \texttt{finalStatePromptJets} and \texttt{chargedFinalStates}.}
\label{tab:pandasCol}
\resizebox{\linewidth}{!}{
\begin{tabular}{ll}
\hline
Column                          & Description                                                                              \\ \hline
eventNumber                     & Unique identification number for each $pp$ collision event.                              \\
particleIndex                   & Unique identification number for the particular particle in an event.                    \\
px, py, pz, pt                  & The $x$, $y$, $z$ and transverse component of the particle momentum respectively in MeV. \\
E                               & The particle's energy in MeV.                                                            \\
mass                            & The particle's mass in MeV.                                                              \\
charge                          & The particle's charge in units of the electron charge.                                   \\
PID                             & The PDG ID of the particle.                                                              \\
status                          & The status of the particle as defined by the MC generator.                               \\
prodVertex, decayVertex &
  \begin{tabular}[c]{@{}l@{}}The 4-position, ($x$,$y$,$z$,$t$), for the production and decay vertex of the given particle in mm and seconds; \\ if the particle has no decay vertex it is set to (-1,-1,-1,-1).\end{tabular} \\
prodVertexDist, decayVertexDist & The radial distance between the production or decay vertex and the IP in mm.             \\
phi, theta, eta &
  The azimuthal angle, $\phi$, the polar angle, $\theta$, and the pseudorapidity, $\eta$ in the \atlas coordinate system. \\
METx, METy, MET                 & The missing momentum in the $x$, $y$ and transverse directions in MeV.                   \\
boost, beta                     & The Lorentz factor, $\gamma$, and the factor $\beta=\frac{v}{c}$.                        \\
ctau                            & The proper lifetime, $c\tau$, for the particle in mm.                                    \\
nParents, nChildren             & The number of parent or children particles the current particle has.                     \\
parentIndices, childrenIndices &
  The indices of the pandas dataframe corresponding to the parents and children of the current particle. \\
weight                          & The generator weight assigned to the current particle.                                   \\ \hline
\end{tabular}%
}
\end{table*}
When the \texttt{selection\_ready} option is enabled, additional operations from the selection stage are performed at ingestion time: optional \(\phi\)-folding is applied before analysis\footnote{As \anubis is only built on the ceiling, to increase statistical precision all down-going tracks can be reflected upward or $\phi$-folded. This requires a factor of 0.5 to be implicitly included in Equation~\eqref{eq:NLLP} to remove the bias. This cannot be applied for all LLP models as the decay topology may need to be preserved, \eg back-to-back jets.}, prompt jets are built, isolation quantities are attached, and the bundle is pruned to the frames required by the selection engine, namely \texttt{LLPs}, \texttt{LLPchildren}, \texttt{finalStatePromptJets} and \texttt{chargedFinalStates}. The full intermediate bundle, LHE files and decayed HepMC files remain available as explicit opt-in artifacts for legacy workflows, debugging or publication archives.
This design shifts the default archival unit from a large event-record file to a lightweight, analysis-ready representation. To preserve provenance, each row keeps the source HepMC filename and size, the stored bundle size and format, the processing stage, the LLP PDG identifier, scan parameters and widths, \madgraph cards and banners, random seeds, and a stable \texttt{run\_hash}. The accessor API can query events, load or export bundles, re-pack older bundles into the selection-ready format, build dashboard payloads and re-materialise generator outputs from the stored cards, seed and scan metadata when a full HepMC representation is required downstream. Transforms enable events to be exported to JSON or human-readable reports without changing the canonical database record.


\subsection{ModelCore}
\label{sec:ModelCore}
This domain offers the user-facing entry point to model parameters and particles through {\textit{SetAnubisInterface}}. Internally, it builds and partially evaluates a parameter tree (using a proxy to the \textit{DataBase} part), preserves LHA formatted metadata, normalises particle information to a common format and can enrich it from a JSON file when needed during the selection procedure. A dedicated \textit{QCDRunner} provides a description of the strong coupling, $\alpha_s(Q)$, and running masses with threshold matching.


For \textit{QCDRunner}, the three loop expansion of $\alpha_s$ is used in the $\overline{MS}$ renormalisation scheme:
\begin{align}
\alpha_s^{(n_f)}(Q) =
\frac{4\pi}{b_0 L}
& \biggl[ 1 - \frac{2 b_1}{b_0^2}\,\frac{\ln{L}}{L}+\nonumber\\
& \frac{4 b_1^2}{b_0^4 L^2}
\left( (\ln{L}-\tfrac12)^2 + \frac{b_2 b_0}{8 b_1^2} - \frac{5}{4} \right)\biggr],
\label{eq:alphas-analyticreal}
\end{align}

where $L=\ln\frac{Q^2}{\Lambda_{n_f}^2}$, $Q$ is an energy scale, $\Lambda_{n_f}$ is the energy-scale where there are $n_f$ active flavours and $b_{0,1,2}$ are the one–, two– and three--loop coefficients of the QCD \(\beta\)-function in the \(\overline{\text{MS}}\) scheme). The implementation of \textit{QCDRunner} is described in \ref{app:QCD}.

\subsection{BranchingRatio}
\label{sec:BranchingRatio}
The \textit{BranchingRatio} section registers decay channels, enforces kinematic thresholds and charge conservation and is able to compute both partial and total decay widths, which can be used to derive the associated branching ratios, as well as the lifetime of LLPs for the current model. Different strategies including analytic UFO expressions, user-supplied Python-based calculators, interpolation from tabulated files, \madgraph-based estimates and \marty-based amplitude generation and evaluation are available within the pipeline.


Combinations of the different methods are also possible. One can calculate decay widths at the quark level automatically using \marty and get to the hadronic level using form factor inputs in Python.

\subsection{MadGraph (Interface)}
\label{sec:Madgraph}
The MadGraph Interface builds job scripts (see Sections~\ref{sec:signalSimulation} and~\ref{sec:Database}) as typed sections and provides editors for `run' and \madspin cards.
These cards are then fed into \madgraph to generate full events. Execution is recommended to be performed inside Docker to provide a controlled environment, and then the \texttt{Events/} outputs are retrieved. 
To control the number of generated events and the random number seed in a batch system, there is a set of card templates for the run cards which can be modified via the editor; these templates are fetched from the \textit{DataBase} layer.


\subsection{Pythia (Interface)}
\label{sec:PythiaInterface}
The Pythia interface generates a \texttt{.cmnd} file that can be used to generate events in \pythia. It supports parameter scans, runs the event generator and writes to \texttt{.lhe} files, \texttt{.hepmc}~\cite{Buckley:2019xhk} files and compact text summaries. A converter reshapes scan outputs into a \madgraph-like folder structure (as shown in Section~\ref{sec:signalSimulation}) with \texttt{run\_NN} and a \texttt{scan\_run\_output.txt} that the database can ingest.


%
In the \texttt{.cmnd} file, all of the sections presented in  Figure~\ref{fig:cmnddecomposition} can be modified using the interface. General Pythia parameters should not be modified without a detailed understanding of the generator.

The interface can add all the decays of a new particle (and its production) using the \textit{BranchingRatio} part into the \texttt{.cmnd} file. It is also possible to change the lifetime of a new particle within \pythia to make it decay; a lifetime must be set for this, as a prompt particle will not decay automatically with \pythia.

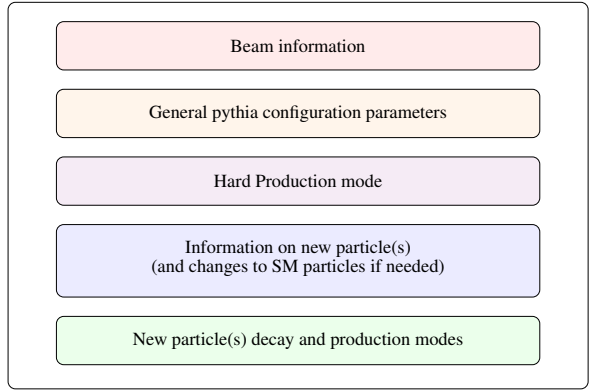
\begin{figure}[ht!]
\begin{center}
\resizebox{\linewidth}{!}{
\begin{tikzpicture}[scale=0.92]
\draw[rounded corners] (-6,2.5) rectangle (6,10.5);
  \draw[rounded corners, fill=green!8] (-5,3) rectangle (5,4);
  \node at (0,3.5) {New particle(s) decay and production modes};
  \draw[rounded corners, fill=blue!8] (-5,4.4) rectangle (5,5.9);
  \node at (0,5.4) {Information on new particle(s)};
  \node at (0,5) {(and changes to SM particles if needed)};
  \draw[rounded corners, fill=violet!8] (-5,6.3) rectangle (5,7.3);
  \node at (0,6.8) {Hard Production mode};
  \draw[rounded corners, fill=orange!8] (-5,7.7) rectangle (5,8.7);
  \node at (0,8.2) {General pythia configuration parameters};
  \draw[rounded corners, fill=red!8] (-5,9.1) rectangle (5,10.1);
  \node at (0,9.6) {Beam information};
\end{tikzpicture}
}
\end{center}
\caption{Structure of the \texttt{.cmnd} file within the \pythia Interface.}
\label{fig:cmnddecomposition}
\end{figure}

The binding with the \pythia{8} software is achieved using pybind11, allowing control within a Python environment and also generator level cuts.

\subsection{Geometry}
\label{sec:geometry}
The \textit{Geometry} section builds a representation of the \atlas cavern geometry and \anubis RPC stations, including vaulted ceilings, service shafts PX14/PX16, and the locations of points of interest such as the interaction point and the centre of curvature of the vaulted ceiling. It exposes coordinate transforms, acceptance predicates and intersection routines. Builders materialise geometries from configuration, with optional caching for repeated runs. Plotting adapters provide standard projections that can be used to verify the topology of LLP events visually. The main use of this section is to apply geometric acceptance requirements when selecting LLP events.


\subsection{Selection}
\label{sec:selection}
This part applies a fixed-order set of selections to generator-level event bundles. The purpose of which is to check whether the LLPs have decayed (\texttt{LLPDecay}); the geometric acceptance of the LLP vertex, \ie whether it is within the \atlas cavern (\texttt{InCavern}), outside the bounds of \atlas (\texttt{NotInATLAS}), and within the active volume of \anubis (\texttt{Geometry}); whether the LLP tracks would intersect with \anubis RPC layers and thus be recorded (\texttt{Tracker}); missing transverse momentum thresholds (\texttt{MET}); and isolation against prompt particles (\texttt{IsoCharged}) or jets (\texttt{IsoJets}) which is combined as \texttt{IsoAll}.  Each of these represents a stage of the cutflow described by the associated names given here. The output of the selection is a dictionary of the number of remaining events compared to the original number, \texttt{Original}, for each of these stages with the number of events after all selections given by \texttt{Final}. This data can be sourced from event bundles, HepMC files or the events database; utilities exist to index and aggregate per-sample cutflows.

There is also the option to re-weight the vertex position based on the lifetime and boost of the LLP, which enables effective resource allocation when there is an effective way to obtain the expected LLP lifetime in a model for a given set of parameters. In that circumstance, events need only be generated for a single set of parameters and then re-weighted post-generation to another arbitrary parameter set. This saves both CPU time and artifact storage space. 


\subsection{Scientific event and campaign inspection}
\label{sec:visualisation-tools}
Two optional Dash applications are distributed with \setanubis as inspection tools rather than as components of the physics calculation. Both consume the same public data structures as scripts and tests; they do not alter geometry definitions, selection thresholds or stored campaign\footnote{A campaign refers to a set of events generated under the same conditions with the same selections applied; this can be tagged within the database.} records.

The \emph{HepMC selection explorer} starts from the compact seven-event HepMC2 reference sample used by scenario R5 (See Section~\ref{sec:RefScenarios}) when no user file is supplied. Its interface follows the canonical selection sequence---\texttt{Original}, \texttt{LLPDecay}, \texttt{InCavern}, \texttt{NotInATLAS}, \texttt{Geometry}, \texttt{Tracker}, \texttt{MET}, \texttt{IsoJets}, \texttt{IsoCharged}, \texttt{IsoAll} and \texttt{Final}---and reports the cumulative cutflow together with the last passed and first failed stage for each event. The event display combines detector-region classification, two- and three-dimensional geometry views, HepMC ancestry and distributions of missing transverse momentum, transverse momentum, pseudorapidity and decay time. Geometric display categories are kept distinct from the production predicates used for the \texttt{InCavern} and \texttt{NotInATLAS} decisions.

The \emph{campaign database inspector} creates an isolated demonstration workspace by default. The workspace contains a schema-compatible SQLite catalogue, a content-addressed copy of the R5 HepMC sample, compact dataframe and selection-manifest artifacts, and representative HNL particle metadata. Users can then select a real \setanubis database and inspect campaign summaries, storage and provenance, generated samples, particle-model records and arbitrary metadata. The demonstration database is read-only from the perspective of campaign maintenance, preventing an exploratory dashboard session in a GUI from modifying a production catalogue.

\section{Validation, examples and inspection tools}
\label{sec:ValidationAndExamples}

\subsection{Executable examples and release validation}
The installed package contains executable examples for each principal domain. They are import-safe and use a shared runtime entry point, so the \setanubis release banner is shown once during direct execution but never as a side effect of importing an example module. Representative examples are listed in Table~\ref{tab:examples}. Machine-readable commands, including \texttt{setanubis-pythia-check {-}{-}json}, suppress the banner and emit clean JSON on standard output.

\begin{table*}[!t]
\centering
\caption{Representative examples distributed with release \setanubistag. Example files are found within \texttt{setanubis/SetAnubis/examples/} followed by the Domain name, \eg \texttt{setanubis/SetAnubis/examples/ModelCore/example\_setanubis\_interface.py}.}
\label{tab:examples}
\begin{tabularx}{\textwidth}{@{}l l X@{}}
\toprule
Domain & Example & Purpose \\
\midrule
ModelCore & {\footnotesize \texttt{example\_setanubis\_interface.py}} & Load a UFO model, inspect particles and parameters, and modify a benchmark value through the public model interface. \vspace{0.1cm}\\
BranchingRatio & {\footnotesize \texttt{example\_BranchingRatioInterface\_hnl.py}} & Evaluate HNL decay information through the branching-ratio interface and report widths, branching fractions and lifetime quantities.\vspace{0.1cm}\\
Pythia & {\footnotesize \texttt{example\_pythia\_cmnd.py}} & Construct a deterministic \pythia command file without requiring the compiled \pythia binding.\vspace{0.1cm}\\
Pythia & {\footnotesize \texttt{example\_pythia\_run.py}} & Run generation when the optional native binding is available and write the corresponding HepMC output.\vspace{0.1cm}\\
MadGraph & {\footnotesize \texttt{example\_madgraph\_interface.py}} & Assemble command, parameter, run, \madspin and \pythia cards through the public interface.\vspace{0.1cm}\\
MadGraph & {\footnotesize \texttt{example\_hepmc\_plots.py}} & Inspect the HepMC output of a generated \madgraph sample.\vspace{0.1cm}\\
Selection & {\footnotesize \texttt{example\_selection\_pipeline.py}} & Execute the named selection stages on a prepared event representation.\vspace{0.1cm}\\
Selection & {\footnotesize \texttt{example\_real\_selection\_trace\_report.py}} & Run the compact real-event benchmark and write JSON and standalone HTML trace reports.\\
\bottomrule
\end{tabularx}
\end{table*}

\subsection{Minimal public-interface examples}
\label{sec:MinimalExamples}
The concise listings below are abridged from the executable examples distributed with the package. They use the supported public API together with the packaged compact reference sample, and are intended to make the principal model, event-generation and selection contracts explicit in the manuscript. The complete examples and the R1--R5 scenarios remain the authoritative executable records.

The first example loads the packaged HNL UFO model, updates the HNL mass parameter and queries the corresponding particle record. The returned mass is evaluated from the same parameter tree used by the card-generation and decay layers.


\begin{lstlisting}[language=python]
from setanubis import SetAnubisInterface, ufo_path

model = SetAnubisInterface(ufo_path("UFO_HNL"))
model.set_leaf_param("mN1", 5.0)  # GeV

hnl = model.get_particle_info(9900012)
mass = model.get_particle_mass(9900012)
print(hnl["name"], mass)
\end{lstlisting}

The following \madgraph example constructs the command and auxiliary cards for an HNL production sample. Card serialisation is deterministic and does not start \madgraph or require Docker; external execution is requested separately through a runner adapter.

\begin{lstlisting}[language=python]
from setanubis import (
    GeneralCardInterface,
    MadGraphCommandConfig,
    SetAnubisInterface,
    ufo_path,
)

model = SetAnubisInterface(ufo_path("UFO_HNL"))
cards = GeneralCardInterface(
    MadGraphCommandConfig(
        neo_set_anubis=model,
        cache=False,
        model_in_madgraph="SM_HeavyN_CKM_AllMasses_LO",
        shower="py8",
        madspin="ON",
    )
)
cards.run_card_builder.set("nevents", 1000)
cards.jobscript_builder.add_process(
    "generate p p > n1 ell"
)
cards.jobscript_builder.set_output_launch(
    "hnl_sample"
)
cards.jobscript_builder.configure_cards()
job_script = cards.jobscript_builder.serialize()
\end{lstlisting}

The selection example below\footnote{Here the setanubis import refers to the package itself and SetAnubis refers to a folder in the git repository.} uses the packaged seven-event HNL bundle and the reference configuration employed by scenario R5. The source adapter, run configuration, cutflow and event-level trace are standard public objects; the helper functions only provide compact versioned inputs and the corresponding geometry configuration.


\begin{lstlisting}[language=python]
from setanubis import EventsBundleSource, RunConfig
from SetAnubis.examples.Selection.compact_sample import (
    build_selection_config,
    build_selection_pipeline,
    load_compact_bundle,
)

source = EventsBundleSource.from_bundle_dict(
    load_compact_bundle()
)
result = build_selection_pipeline().run(
    source,
    build_selection_config(),
    RunConfig(capture_intermediate=True),
)

cutflow = result["cutFlow"]
print(cutflow["nLLP_original"], 
      cutflow["nLLP_Final"])
print(result["trace"].event_summary[
        ["eventNumber", "last_passed_stage"]
    ]
\end{lstlisting}

The complete dry-run \madgraph example and the selection-trace report can be executed directly after installation. The first command writes the generated card contents to standard output without launching an external generator; the second writes machine-readable JSON and a standalone HTML report.
\begin{lstlisting}[language=bash]
python -m SetAnubis.examples.MadGraph.example_madgraph_interface
python -m SetAnubis.examples.Selection.example_real_selection_trace_report \ 
       --output-dir selection_trace_output
\end{lstlisting}

Release \setanubistag is tested on Python 3.10--3.13. The release snapshot contains 247 automated unit, integration, packaging and contract tests and reaches 67\% line coverage for the measured Python modules, above the release threshold of 58\%. These software tests verify interfaces, deterministic serialisation, database migrations, geometry predicates, cutflow bookkeeping, distributable resources and installation from the wheel; they complement, rather than replace, model-specific physics validation. Continuous integration also applies Ruff static checks, a high-severity Bandit scan, strict Sphinx documentation builds, package construction and \texttt{twine check}, and executes the R1--R5 reproducibility suite before publication.

\subsection{Heavy Neutral Lepton benchmark}
\label{sec:ToyCases}

As a benchmark for the branching-ratio interfaces, a HNL model is considered. This same model was previously studied with an earlier prototype of the framework to perform a dedicated \anubis sensitivity study with \setanubis~\cite{Reymermier:2026HNL}. 
The Lagrangian of this model is the following:
\begin{equation}
    \mathcal{L}_{\text{HNL}} = \frac{i}{2}\overline{N_1} \slashed{\partial}N_1 -\frac{m_{N_1}}{2} \overline{N_1} N_1^c-y_{i \alpha} \overline{N_1} \tilde{\phi}^\dagger L^\alpha + h.c.,
    \label{eq:HNL_Lagrangian}
\end{equation}
where $\alpha \in \{e,\mu,\tau\}$, $N_1$ is a single right-handed HNL field with the mass $m_{Ni}$, and $y_{i\alpha}$ is a matrix of complex Yukawa couplings~\cite{Li_2023,Brdar_2019}. 
Here, $L^\alpha$ is the SM left-handed leptonic doublet field, $\phi$ is the Higgs doublet field with $\tilde{\phi} = i\sigma_2 \phi^{*}$ for the second Pauli matrix $\sigma_2$, and $N_i^c = \gamma^2 N_i^{*}$ with $\gamma^2$ the second Dirac matrix.
Taking the interaction terms with the $H$, $W$ and $Z$ bosons one obtains:
\begin{align}
\mathcal{L}_{\text{HNL}}^W &=-\frac{g V_{1\alpha}}{\sqrt{2}} \left[ W^-_\mu(\overline{l}_{\alpha}^L \gamma^\mu N_1)\right]+h.c., \label{eq:WInteractions} \\
\mathcal{L}_{\text{HNL}}^Z&=-\frac{\sqrt{g^2 + g'^2}}{2}V_{1\alpha} Z_\mu\left(\overline{\nu}_{\alpha}^L \gamma^\mu N_1\right)+h.c., \label{eq:ZInteractions}\\
\mathcal{L}_{\text{HNL}}^H &= -V_{1\alpha} \frac{m_{N_1}}{v}H\left(\overline{\nu}^L_{\alpha}N_{1}\right)+ h.c.\,.
\label{eq:HInteractions}
\end{align}
These terms appear from the mixing of the HNL with the left-handed neutrinos. Here $g$ and $g'$ are the constant coupling of the ${SU(2)}_L$ and the $U(1)_Y$ gauge groups, $\gamma_\mu$ are the Dirac gamma matrices, $V_{1\alpha}$ is the mixing coupling between the HNL and the left-handed neutrino $\nu_\alpha$, $l_\alpha^L$ is a left-handed lepton of flavour $\alpha$, and $m_N$ is the mass of the HNL~\cite{Li_2023}.

For the 2-body decays at the partonic level (both to produce HNLs and their decays), the expression within the UFO, which is parsed in the Pipeline, is used and partially evaluated with the SM parameters using PDG values, and then can be expressed as a function of the BSM parameters and evaluated. However, through the use of \madspin alongside \madgraph 3-body decays are also possible to be simulated, though an alternative approach is required for \pythia events.

For the 3-body decays at the partonic level, it is possible to use the \marty Framework, as presented in Figure~\ref{fig:marty_plot_HNL}. \marty calculates all the scattering matrices using Feynman diagrams, and integrates over the phase space to get the decay width. It can also calculate cross-sections for $2\to 2$ or $2\to 3$ kinematics. Figure~\ref{fig:marty_plot_HNL} shows that \marty aligns with the analytical calculations from theory for a HNL that only couples to muons. The $N_1\rightarrow\nu_\mu\mu^+\mu^-$ mode has a larger decay width than the other lepton flavours as it can proceed via charged-current or neutral-current decays, and not just neutral-current as with the $e$ and $\tau$ cases. The small difference between $e$ and $\tau$ arise due to the available kinematic phase space due to the differing masses.
\begin{figure}[ht]
  \centering
  \includegraphics[width=0.98\linewidth]{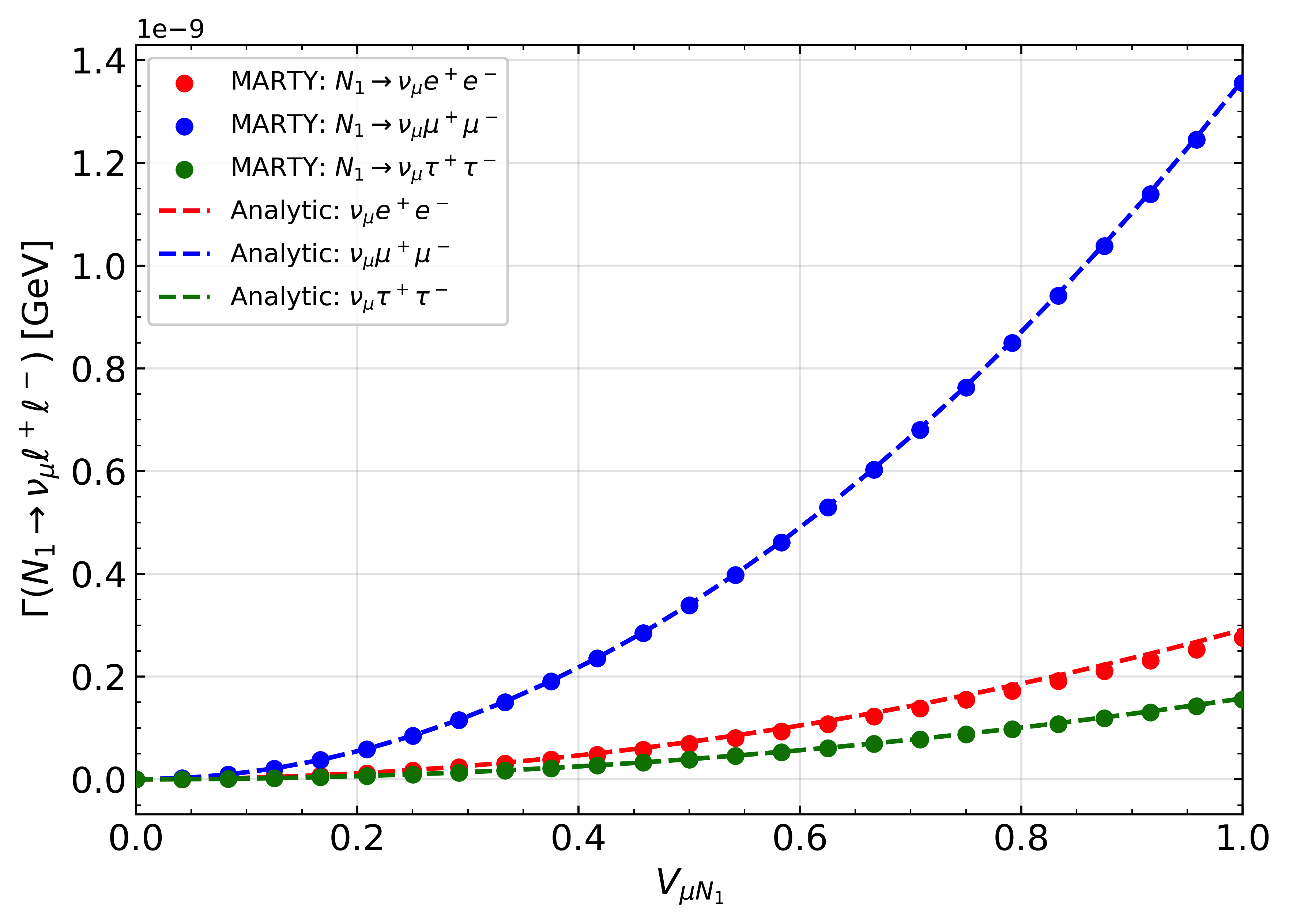}%
  \caption{Example of the Decay Width calculation of LLP (HNL) using \marty for three-body decays compared to analytical curves taken from Ref.~\cite{Bondarenko_2018}. Here the HNL has a mass of 10 GeV and decays into one neutrino ($\nu_\mu$) and a pair of leptons.}
  \label{fig:marty_plot_HNL}
\end{figure}

Decays at the hadronic level are not currently implemented in the pipeline. However, one could use \marty to extract the primary current by isolating the structure:

\begin{equation}
    \mathcal{M}_{\mu}^{\text{lept,BSM}} \times J^{\mu}_{\text{quarks}},
\end{equation}

where $J^{\mu}_{\text{quarks}} = \bar{q}\Gamma^{\mu}q'$ and $\mathcal{M}_{\mu}^{\mathrm{lept,BSM}}$ denotes the leptonic and BSM part of the amplitude, including the relevant couplings, propagators and external fermion wave functions, with the Lorentz index, $\mu$, left open for contraction with the quark current.
If one considers, for example, the process ${N_1 \to \pi\pi\nu}$,  $J^{\mu}_{\text{quarks}}$ is replaced with its element in the hadronic matrix $\bra{\pi(p_1)\pi(p_2)}\bar{q}\Gamma^\mu q' \ket{0}$. A vector current related to the $\pi^+\pi^-$ current is given by $J_V^{\mu} = \bar{u}\gamma^\mu u-\bar{d}\gamma^\mu d$, where $\Gamma^\mu$ denotes the Dirac and chiral structure of the interaction. This gives:
\begin{equation}
    \bra{\pi^+(p_1)\pi^-(p_2)}J_V^\mu \ket{0} = (p_1-p_2)^\mu F_V^{\pi\pi}(s)
\end{equation}
where $s$ is the centre of mass energy and $F_V^{\pi\pi}$ is a vectorial form factor which can be taken from experimental fits, results from lattice QCD or light cone sum rules. This needs to be replaced in the matrix element, $\mathcal{M}$, of \marty and integrated over the phase space, which could be added to the framework in a future version.

To replicate Ref.~\cite{Reymermier:2026HNL}, one can adapt the \madgraph example from Section~\ref{sec:MinimalExamples}. There were three main production modes for the HNL used: Charged and Neutral Current Drell Yan (CCDY and NCDY), and $W\gamma$ fusion. The appropriate processes for each are given below. 
\begin{lstlisting}[language=python]
    # CCDY
    cards.jobscript_builder.add_process(
        "generate p p > n1 ell" ) 
    # NCDY
    cards.jobscript_builder.add_process(
        "generate p p > n1 vv" )
    # WGamma
    cards.jobscript_builder.add_process(
        "generate q a > n1 ell q QED=3 QCD=0" )
    cards.jobscript_builder.add_process(
        "add process a q > n1 ell q QED=3 QCD=0")
\end{lstlisting}
Here \texttt{ell} and \texttt{vv} refer to being $e^+/e^-$ and $\bar{\nu}_e, \nu_e$ respectively, as this example is for BC6 where the HNL couples only to an $e$. To add alternative defined parameters in the \madgraph card, \ie for BC7 where HNL couples only to the $\mu$, this can be done via the following command.
\begin{lstlisting}[language=python]
    cards.jobscript_builder.add_define(
        "mull",["mu+","mu-"])
\end{lstlisting}
There were four possible decays of the HNL, $(N_1\rightarrow X)$, considered in Ref.~\cite{Reymermier:2026HNL} where: $X=q\bar{q}'\ell$, $X=q\bar{q}\nu_{\ell}$, $X=\ell\bar{k}\nu_k$, or $X=k\bar{k}\nu_{\ell}$, with $\ell=e$ or $\mu$, ${k=e,\mu}$ or $\tau$. For these, the \madspin card is modified as below, where the last two decays can be combined as the HNL is set only to decay to a single flavour in \madgraph.
\begin{lstlisting}[language=python]
    cards.madspin_builder.add_decay(
       "decay n1 > q q ell")
    cards.madspin_builder.add_decay(
       "decay n1 > q q vv")
    cards.jobscript_builder.add_define(
       "lep",["e+","e-","mu+","mu-","ta+","ta-"])
    cards.jobscript_builder.add_define(
       "vlep",["ve","ve~","vm","vm~","vt","vt~"])
    cards.madspin_builder.add_decay(
       "decay n1 > lep lep vlep")
\end{lstlisting}
Finally, the card needs to be set up for the parameter scans in terms of the HNL mass, MN1, and the coupling, VeN1, for the chosen values of each. 
\begin{lstlisting}[language=python]
    job_card.add_parameter_scan(
        "VeN1", "[1e-6, 1.]")
    job_card.add_parameter_scan(
        "MN1", "[0.5, 1.0]")
\end{lstlisting}
Following this, a set of \madgraph samples equivalent to those used in Ref.~\cite{Reymermier:2026HNL} can be produced. Then, if one uses the selection example from Section~\ref{sec:MinimalExamples}, the default selection pipeline matches that used for the \anubis HNL sensitivity estimate. From which the cutflows can be extracted for each \madgraph run and from there processed into plots over the whole run as shown in Figure~\ref{fig:cutFlow} or used with the DataBase information to evaluate Equation~\eqref{eq:NLLP} and find the sensitivity contours, like those shown in Figure~\ref{fig:SenContours}. 

\begin{figure}[htb]
    \centering
    \includegraphics[width=\linewidth]{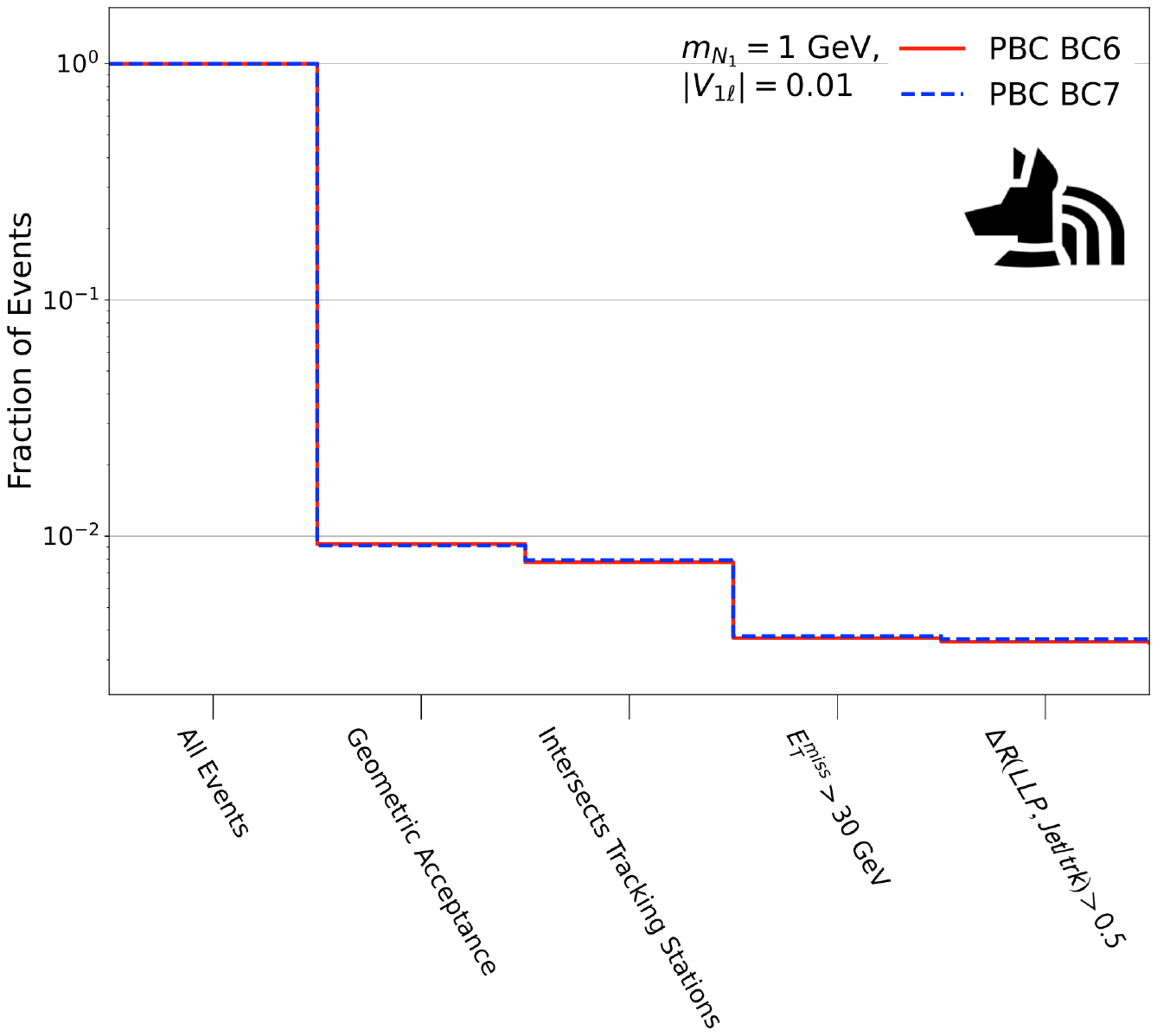}
    \caption{The cutflow for BC6 and BC7, where $m_{N_1}=1~\GeV$ and $|V_{1\ell}|=0.01$ produced from applying the selection pipeline. Figure taken from Ref.~\cite{Reymermier:2026HNL}.}
    \label{fig:cutFlow}
\end{figure}

\begin{figure}[htb]
    \centering
    \includegraphics[width=0.9\linewidth]{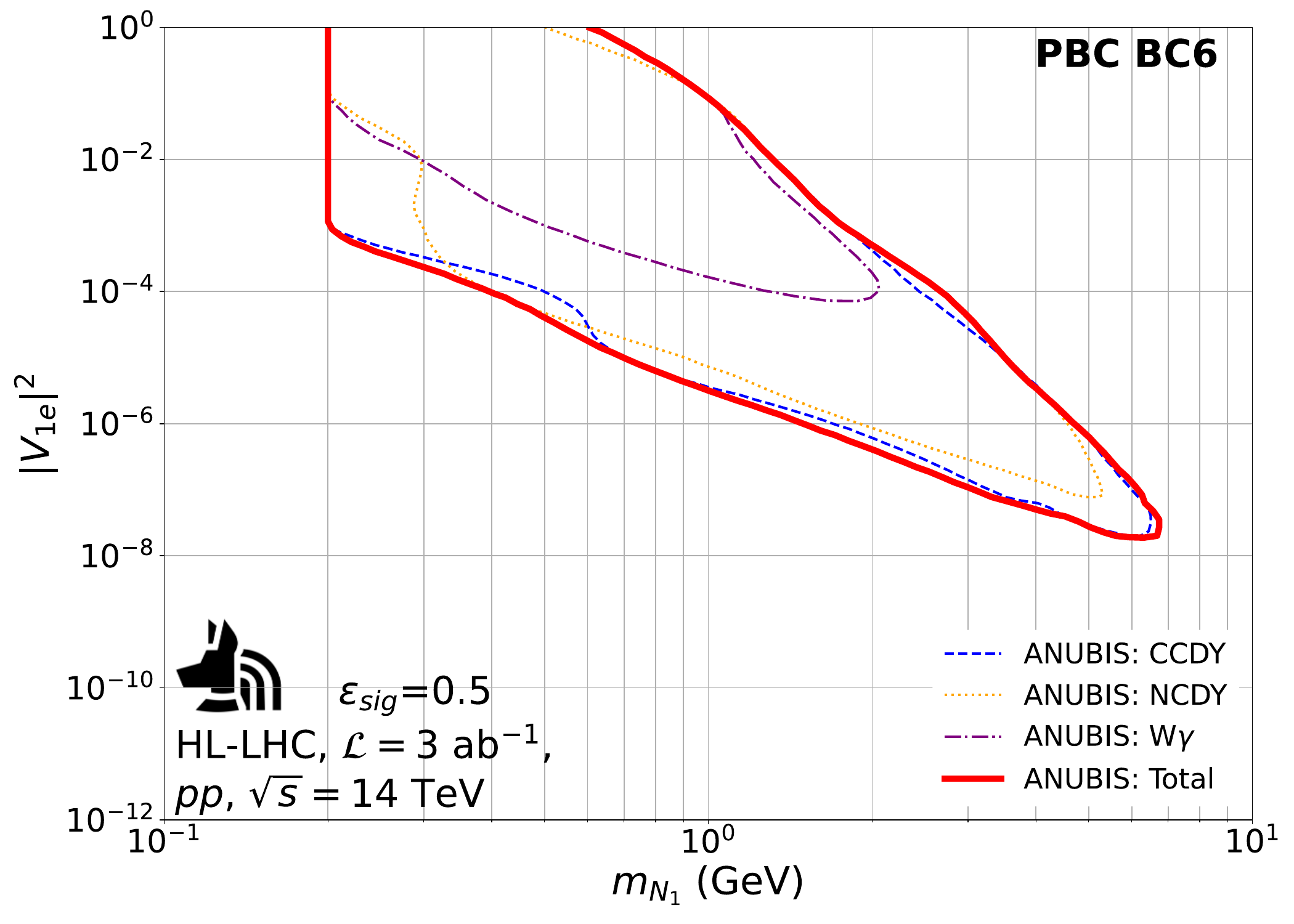}
    \caption{Example contour produced from evaluation of the cutflows and information from the DataBase for BC6, showing the relative contributions of the three HNL production modes. Figure taken from Ref.~\cite{Reymermier:2026HNL}.}
    \label{fig:SenContours}
\end{figure}


\subsection{Geometric acceptance and RPC intersections}
\label{sec:GeometricAcceptance}


The geometry implementation was validated independently of the event-generation chain using a Monte Carlo volume-reconstruction test. Probe points were sampled uniformly in bounding boxes enclosing the relevant cavern and detector regions. For each probe point, the same public geometry predicates used by the selection API were evaluated to determine whether the point lay inside the \atlas cavern volume, outside the \atlas detector envelope, within one of the service shafts, or inside an \anubis RPC module. The corresponding volume estimate was then obtained from
\begin{equation}
    \widehat{V}_{\mathrm{MC}} = V_{\mathrm{box}}\,\frac{N_{\mathrm{accepted}}}{N_{\mathrm{sampled}}},
\end{equation}
with the binomial Monte Carlo uncertainty propagated from the accepted fraction. The reconstructed volumes were compared with the reference volumes computed directly from the geometry definition. Within the statistical precision of the sampling, the MC estimates reproduced the expected cavern, \atlas and RPC volumes, providing a check of the coordinate transforms and region-classification predicates.

A complementary track-based test was performed for the RPC-intersection routines. Straight LLP and daughter-particle trajectories were generated with random starting points and directions, and the same intersection APIs used in the selection cutflow were used to record crossings with the RPC layers. The resulting hit patterns were checked against the known position and orientation of the RPC planes and were inspected with the geometry visualisation tools in the $xy$, $xz$, $zy$ and three-dimensional projections. These tests verify that the geometric acceptance used in the cutflow is consistent with the implemented detector volume and that boundary cases near the cavern walls, shafts and RPC surfaces are handled coherently and appropriately.

\section{Reproducibility and availability}
\label{sec:Reproducibility}

\subsection{Reference scenarios R1--R5}
\label{sec:RefScenarios}
The release repository contains five independent experiments under \texttt{reproducibility/}, summarised in Table~\ref{tab:repro-scenarios}. Each experiment has an \texttt{input/} directory with versioned parameters or a reference to a packaged resource, an \texttt{expected\_output/} directory with a compact versioned summary, an ignored \texttt{output/} directory for generated results, a standalone \texttt{run.py}, and a README describing the scientific scope. Generated outputs are deliberately excluded from Git: they are reconstructed locally or in continuous integration and can be archived as release evidence without mixing transient products into the source history.

\begin{table*}[t]
\centering
\caption{Reproducibility scenarios included in \setanubistag.}
\label{tab:repro-scenarios}
\begin{tabularx}{\textwidth}{@{}l l X X@{}}
\toprule
ID & Domain & Input and operation & Validated output \\
\midrule
R1 & Core/model & Load the packaged HNL UFO model, inspect its particle/parameter content and modify the \texttt{mN1} benchmark. & Deterministic model identity, particle counts and updated parameter value.\vspace{0.1cm}\\
R2 & Branching ratio & Evaluate tabulated/interpolated partial widths for a fixed HNL point. & Partial widths, total width, normalised branching ratios and lifetime summary.\vspace{0.1cm}\\
R3 & Pythia & Build the \pythia configuration from versioned parameters without starting \pythia. & Byte-stable \texttt{.cmnd} content and a structured summary of the configured process.\vspace{0.1cm}\\
R4 & MadGraph & Build the command, run, parameter, \pythia{8} and \madspin cards without starting \madgraph. & Deterministic card digests and selected physical/run settings.\vspace{0.1cm}\\
R5 & Selection & Read the packaged compact HepMC2 benchmark, reconstruct selection dataframes and execute the canonical cutflow. & Cutflow summary, reconstructed event bundle, JSON trace, standalone HTML trace and validation marker. \\
\bottomrule
\end{tabularx}
\end{table*}

The complete suite is run from a clean checkout with
\begin{lstlisting}[language=bash]
python -m pip install -e ".[dev,selection]"
python reproducibility/run_reproducibility.py \
  --output-root reproducibility_outputs
\end{lstlisting}
A single scenario can be selected with {\texttt{{-}{-}scenario R5}}. The same command is a required GitHub Actions gate and its output is retained as a workflow artifact. R3 and R4 isolate deterministic preparation from external generator execution, so the reference suite remains runnable without \madgraph, \marty or a compiled \pythia binding.

\subsection{Selection trace and reference cutflow}
Scenario R5 uses a seven-event HepMC2 sample in which events are constructed to fail at different selection stages. The expected cumulative counts are 7 at \texttt{Original} and \texttt{LLPDecay}, 6 at \texttt{InCavern}, 5 at \texttt{NotInATLAS}, 4 at \texttt{Geometry}, 3 at \texttt{Tracker}, 2 at \texttt{MET}, and 1 after each isolation stage and at \texttt{Final}. This compact sample is not intended to represent a physical signal distribution; it is a transparent regression fixture for the interpretation and ordering of the production cuts.

The trace writer produces two files called \texttt{selection\_trace} but with different file extensions,  \texttt{.json} and \texttt{.html}. The JSON document contains configuration provenance, transition counts and per-event decisions. The HTML report embeds its styles and data
and presents the cumulative cutflow together with the last successful and first failed stage of every event. It is therefore suitable for human review, supplementary release material and long-term audit while retaining the JSON file as the machine-readable source of truth.

\subsection{Campaign provenance and content-addressed storage}
Generation runs are indexed by stable hashes computed from their relevant cards, banners, scan metadata and event records. Heavy artifacts are stored once in a content-addressed store keyed by SHA-256, while an SQLite catalogue in WAL mode records models, events, scan coordinates, bundle formats and processing history. The default persistent event representation is a compact selection-ready dataframe bundle; retaining raw HepMC is an explicit policy choice for benchmarks or analyses requiring the full event record. Campaign identifiers are derived from the sorted run hashes and a serialised manifest of fixed settings, allowing an imported campaign to be checked or re-materialised without relying on directory names.

\subsection{Software, documentation and archived release}
The development repository is \setanubisrepo, and the immutable source state described in this article is identified by tag \setanubistag. The user and API documentation is published at \setanubisdocs. For the final tagged release, the automated publication workflow first uploads the wheel and source distribution to TestPyPI (\setanubistestpypiurl), downloads the wheel again, verifies its SHA-256 checksum and executes installation smoke tests. The unchanged artifacts are then promoted to PyPI (\setanubispypiurl) as \setanubispypi. TestPyPI is therefore a staging and verification channel, while PyPI is the supported user-facing package index. The archival Zenodo identifier for the final source release is \setanubisdoi~\cite{bib:SETANUBISSoftware}.

\section{Computational performance}
\label{sec:Performance}
\subsection{Card construction and execution time}
\label{sec:CardConstruction}
Card construction is deterministic and fast compared to event generation. For \pythia, configuration files are produced as ordered sections and serialised linearly; the cost grows with the number of declared particles and decays but remains essentially negligible once templates are in place. For \madversion{3.5.8}, jobscripts and card editors operate on text; parsing and re-serialisation dominate when many comments or aligned columns are preserved, yet the wall-clock time is still small compared to a typical run.

The build time, $T_{\mathrm{build}}$, is modelled as
\begin{align}
    T_{\mathrm{build}} \simeq a_0 &+ a_1\,N_{\text{sections}} + a_2\,N_{\text{particles}}\\
    &+ a_3\,N_{\text{decays}} + a_4\,N_{\text{scan points}}.\nonumber
    \label{eq:buildTime}
\end{align}
Here $N_{sections}$ denotes the number of configuration blocks to be emitted, $N_{particles}$ and $N_{decays}$ count the model entries that must be materialised in the cards, and $N_{scan points}$ is the number of parameter points for which cards are instantiated. The constants encapsulate template I/O and in-memory editing. In practice, \(T_{\mathrm{build}}\) is amortised over scans and cached artifacts.

For a campaign consisting of \(R\) runs with \(N_r\) events each the execution time, $T_{\mathrm{exec}}$ , can be modelled as,
\begin{equation}
    T_{\mathrm{exec}} \simeq \sum_{r=1}^{R}\big(T_{\mathrm{init}}^{(r)} + N_r\,t_{\mathrm{event}}^{(r)}\big),
    \label{eq:ExecutionTime}
\end{equation}
where \(t_{\mathrm{event}}\) is the measured time per generated event (this is different for parton-level generation vs showering/hadronisation, and if there are selections at the matrix-element level, or for high-multiplicity final states). In practice the execution time dominates the other sources. When \madgraph is containerised, its initialisation time, $T_{\mathrm{init}}^{(r)}$, is
\begin{equation}
    T_{\mathrm{init}}^{(r)} = T_{\text{container start}} + T_{\text{generator init}} + T_{\mathrm{I/O}},
\end{equation}
with a one-off image pull on first use. For large \(N_r\), the linear term dominates and the container overhead is negligible; for small ``pilot'' runs, the opposite holds.

Throughput should be characterised empirically on the target machine and image. A short pilot (\(N_r\) in the \(10^3\!-\!10^4\) range) provides \(t_{\mathrm{event}}\) and an uncertainty; those numbers then forecast the wall-clock of the full scan. To ensure that throughput measurements are reproducible and comparable across machines, the benchmarking metadata should include the CPU model, core count, SMT configuration, kernel version, Docker image digest, generator versions, and the random seeds used. A cached version of the Geometry and database WAL mode reduces secondary costs (geometry rebuilds, small transactional writes) without changing physics results.

\subsection{Direct-HepMC selection throughput}
\label{sec:SelectionScaling}

To characterise the computational cost of the event-selection stage itself, a benchmark was performed starting from a pre-generated HepMC sample and excluding the event-generation step. The benchmark therefore measures the complete \setanubis processing chain
for an HNL benchmark sample. The benchmark was executed for increasing event counts, and the measured wall-clock times are shown in Fig.~\ref{fig:selection_time_scaling_legacy}.

Over the tested range, the execution time grows approximately linearly with the number of input events, indicating that the direct-HepMC (without storage inside the database) selection workflow is well described by an approximately constant effective processing time per event. In the tested range shown in Fig.~\ref{fig:selection_time_scaling_legacy}, the measured cost is of order $70$--$80~\mathrm{ms}$ per event, with the total wall-clock time reaching about $287~\mathrm{s}$ for $3600$ input events. This benchmark includes not only the final cutflow evaluation, but also the HepMC ingestion, dataframe construction, LLP-bundle preparation, and the jet/isolation preprocessing steps required before the selection is applied.

\begin{figure}[t]
    \centering
    \includegraphics[width=\linewidth]{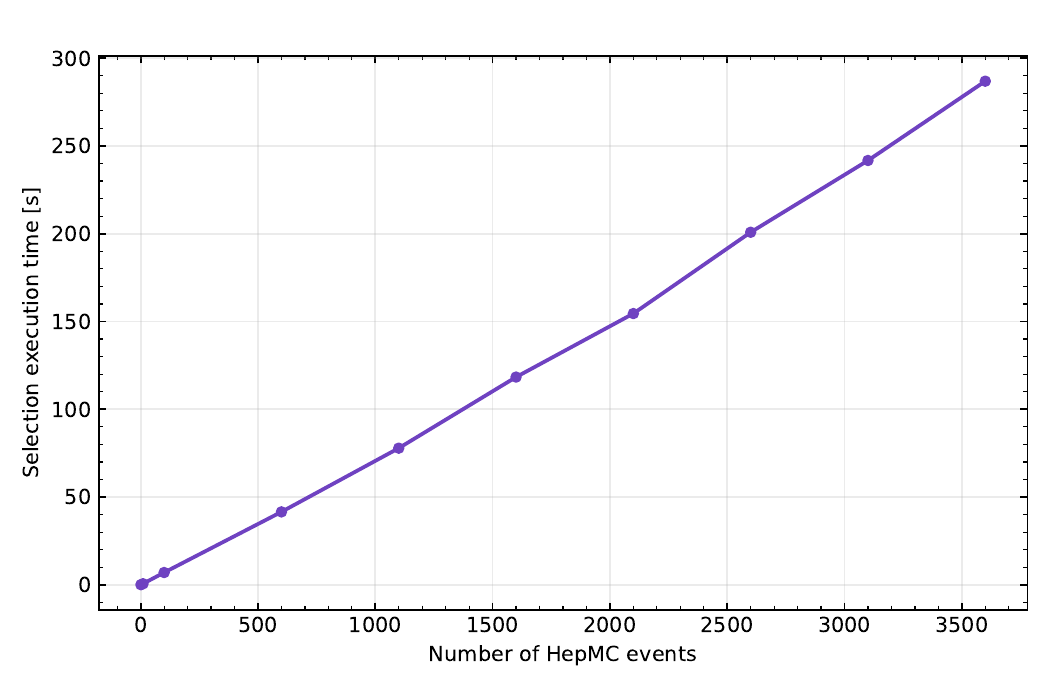}
    \caption{Wall-clock time of the direct-HepMC selection benchmark as a function of the number of input events. The benchmark includes HepMC ingestion, dataframe construction, LLP-bundle preparation, jet/isolation precomputation and the full selection sequence, but excludes event generation.}
    \label{fig:selection_time_scaling_legacy}
\end{figure}

All benchmark measurements reported in this section were performed on a workstation equipped with an Intel\textsuperscript{\textregistered} Core\textsuperscript{\texttrademark} i9-14900K processor (3.20~GHz) and 64~GB of RAM (63.7~GB usable), running Ubuntu 24.04 through Windows Subsystem for Linux (WSL).

\subsection{Scan cost and dataframe-bundle storage}
\label{sec:ScanCost}
Let a scan span \(d\) parameters with grid sizes \(k_1,\dots,k_d\). The number of runs is \(R=\prod_i k_i\). For a fixed per-point sample size, \(N\), the statistical relative uncertainty on an efficiency \(\hat{\varepsilon}\) estimated from accepted events obeys
\begin{equation}
    \frac{\sigma_{\hat{\varepsilon}}}{\hat{\varepsilon}}
= \sqrt{\frac{1-\hat{\varepsilon}}{N\,\hat{\varepsilon}}}\,.
\end{equation}
Solving for \(N\) gives \(N \simeq (1-\hat{\varepsilon})/(\hat{\varepsilon}\,\delta^2)\) to reach a target relative error, \(\delta\). For very small acceptances the denominator is \(\hat{\varepsilon}\), so \(N\) must grow accordingly; for feasibility studies, it is common to stage the scan with a light pilot to map where \(\hat{\varepsilon}\) is non-zero and then generate a large number of events.

With the current database backend, storage can be described in terms of the compact derived bundle rather than the raw event file:
\begin{equation}
    S_{\mathrm{total}} \simeq \sum_{r=1}^{R}\left(S_{\mathrm{bundle}}^{(r)} + S_{\mathrm{meta}}^{(r)} + I_{\mathrm{HepMC}}^{(r)}S_{\mathrm{HepMC}}^{(r)}\right) + S_{\mathrm{indices}},
\end{equation}
where \(S_{\mathrm{bundle}}^{(r)}\) is the compressed dataframe bundle stored for run \(r\), \(S_{\mathrm{meta}}^{(r)}\) contains cards, banners, scan information and JSON metadata, and \(I_{\mathrm{HepMC}}^{(r)}\) is one only when raw HepMC retention has been explicitly requested. In the default mode, \(I_{\mathrm{HepMC}}^{(r)}=0\): the source HepMC file is read during ingestion, but the persistent artifact is the selection-ready bundle.

The bundle size depends on the number of accepted objects retained per event and on the selected processing stage. A raw \texttt{LLPAnalyzer} bundle stores more intermediate frames and is useful for debugging; a selection-ready bundle precomputes prompt jets and isolation and keeps only the frames required by the selection engine. The database records both the source HepMC size and the stored-bundle size, so storage savings can be monitored per event, per model and per campaign. This is important because large scans are usually limited not only by CPU time but also by transfer, archival and query costs.

In practice, the overall computational cost is governed by four variables: the grid size \(R\), the per-point statistics \(N\), the generator throughput, and the chosen event representation. A pragmatic strategy is to (1) profile \(t_{\mathrm{event}}\) and the post-ingestion bundle size with a pilot on the production image, (2) pick \(N\) from the uncertainty target using the binomial relation above, (3) store selection-ready bundles by default, and (4) retain raw HepMC only for benchmark points, publication checks or workflows that explicitly require the full event record. With those inputs, \(T_{\mathrm{exec}}\) and \(S_{\mathrm{total}}\) follow directly from the formulas above and can be reported alongside the campaign manifest for full reproducibility.

Additional improvements could be made by using the lifetime re-weighting capabilities, as often the kinematics of the LLP decay are independent of the decay position. In such circumstances, if a relationship exists between the lifetime of the LLP and a set of scan parameters, then a single set of simulations in terms of parameters that the LLP lifetime does not depend on could be reused with a random decay position. The unweighted samples can be cached before lifetime re-weighting and after the most computationally costly aspects, such that additional re-weighted samples can share a common base file to produce the final output artifacts with much reduced wall-clock time. If there is no clearly defined relationship between the scan parameters and the lifetime, these values could be extracted by performing a grid scan with a low-level set of pilot samples to extract an empirical lifetime for each grid scan point in use and use that to perform the re-weighting.


\section{Limitations and roadmap}
\label{sec:Limitations}
\subsection{Detector response simulation}
\label{sec:Limit-DetResponse}
At present, \setanubis is limited to particle-level simulation of the production and decay of the given physics model, and then projecting those tracks to a simplified geometric model. There is no consideration of simulating the detector response, hit smearing, time or spatial resolution of the detector, the influence of a magnetic field or material interactions. These are all aspects that would need to be considered for the real detector, which have currently been simplified by using truth positions and introducing an optional efficiency for the RPCs. There are plans to introduce the option to use a full \geant~\cite{GEANT4:2002zbu} simulation of the \anubis detector that is being developed. Due to the modular nature of the framework, this could be introduced as part of the selection without requiring significant additional effort. 

\subsection{Hadronic decays and form factors}
\label{sec:Limit-HadronicFF}
\setanubis can compute widths and branching ratios through several strategies, but hadronic channels remain a limiting factor when reliable form factors are unavailable or when the relevant degrees of freedom are hadronic rather than partonic. Analytic UFO functions typically cover two-body decays with point-like couplings; multi-body decays and channels mediated by hadronic resonances are either absent or require model-specific inputs. In such cases the framework falls back to user-provided calculators or tabulated rates. This provides practical coverage but leaves an intrinsic modelling uncertainty that must be reported with the results.

\marty-based amplitudes help when analytic expressions are feasible, yet they require a dedicated model file and a stable mapping between UFO and \marty conventions. For the moment, one needs to create a model file for any BSM extension before using \marty. The current bridge handles tree-level and selected one-loop structures for two-body and three-body decay (and $2\rightarrow2$, $2\rightarrow3$ cross sections); it does not guarantee full coverage of hadronic matrix elements or model-specific hadronisation assumptions. Numerical stability can also degrade near thresholds or narrow-width poles unless kinematic regularisation is applied consistently across the chain.

The roadmap is to integrate external form-factor inputs for common hadronic final states and to allow per-channel substitution of validated semi-empirical formulae where first-principles amplitudes are not yet robust. On the \marty side, there are plans to cache code generation artifacts, expand the UFO$\rightarrow$\marty mapping tests, and expose a small validation suite that compares partial widths against reference values (analytic or experimental) before a campaign. 
Additionally, there are plans to add improved plotting tools to produce common sets of sensitivity plots at the end of the pipeline, though these may require some user intervention to suit the targeted models. These changes do not alter the domain APIs; they provide better defaults and tighter checks while keeping the adapter boundaries intact.

\subsection{Storage optimisations and parallelisation}
\label{sec:StorageOptimisations}

Large event campaigns remain constrained by both storage and wall-clock cost, even after the default storage model is changed from raw HepMC to compact dataframe bundles. The current approach reduces the persistent footprint by storing only the analysis-ready representation in the CAS while retaining the source-file sizes and generation metadata needed for audit. Further gains are expected from more systematic pruning of unused columns, optional columnar formats for purely numerical frames, and per-geometry caches keyed by configuration so that repeated scans do not rebuild identical detector objects.

Parallel execution is most effective at the scan level, with independent runs distributed across cores or nodes. Containerised \madgraph isolates toolchains and avoids version drift; however, launching too many containers saturates I/O and reduces throughput. A practical policy is to cap concurrent jobs to the number of physical cores or the I/O bandwidth of the target system, pin random seeds per run for reproducibility, and stage scans so that geometry caches and database writes are reused rather than contended. The database importer benefits from batched transactions under WAL mode; small indices and JSON summaries can be streamed while bundle blobs are written sequentially to the CAS to minimise seek overheads.

Planned improvements include a first-class policy for deciding which benchmark runs should keep raw HepMC, an optional binary or columnar path for selected event summaries, stronger bundle-schema versioning, per-geometry versioned caches keyed by configuration, and a lightweight scheduler that coordinates container concurrency with database ingestion. Together these changes reduce storage, improve I/O locality and make large parameter scans predictable and reproducible on shared systems.

\section{Conclusion}
\label{sec:conclusion}
\setanubis provides an auditable path from model parameters and decay information to generated event records, \atlas/\anubis geometry and a named LLP-selection cutflow. Its ports-and-adapters architecture separates physics-domain decisions from generators, storage and visualisation, while compact event bundles and content-addressed campaign records support parameter scans without discarding provenance. Release \setanubistag adds a versioned R1--R5 reproducibility contract, machine-readable and standalone HTML selection traces, scientific event/campaign inspection applications and tested distributions for Python 3.10--3.13. These facilities establish a stable software baseline for the \anubis sensitivity studies~\cite{ANUBIS:2025sgg,Reymermier:2026HNL}. Future work for \setanubis includes broader validation of hadronic decay inputs, increased coverage of external generator execution, further optimisation of large distributed campaigns and introducing simulation of the detector-level response to the pipeline.

\section*{CRediT authorship contribution statement}
\textbf{Anna Mullin:} Methodology, Software (first implementation of the event-selection).
{\textbf{ANUBIS Collaboration:}} provision of computing resources, review of the documents and code, endorsement of SET-ANUBIS as default sensitivity evaluation framework.
\textbf{Paul Swallow:} Methodology, Software (second, extended implementation of the event-selection), Validation, Supervision, Writing -- original draft, review \& editing.
\textbf{Sofie Nordahl Erner:} Conceptualisation, Methodology, Software (initial architectural design and initial implementation of the branching-ratio and decay-width calculation components).
\textbf{Th\'eo Reymermier:} Conceptualisation, Methodology, Software, Validation, Visualisation, Project administration, Writing -- original draft, review \& editing (design and implementation of software architecture; development of production code, automated test suite, public interfaces, graphical user interfaces, release infrastructure and reproducibility workflow; preparation of the manuscript).

\section*{Funding}
This research is funded through the UKRI under the Future Leaders Fellowship scheme (Grant number G107408).

\section*{Data and code availability}
The source code, tests, examples and reproducibility inputs are available at {\setanubisrepo} under the GNU GPL version 3 or later. The version described in this article is tag \setanubistag, and its documentation is available at \setanubisdocs. The release artifacts were rehearsed and verified on TestPyPI (\setanubistestpypiurl) before the identical distributions were published on PyPI (\setanubispypiurl) as \setanubispypi. The five reference scenarios and their expected summaries are distributed in the source archive; generated \texttt{output/} directories are not committed and are recreated by the documented runner. The R5 compact HepMC2 sample, JSON selection trace and standalone HTML report are included as reproducibility resources. The version-specific software archive will be deposited on Zenodo at \setanubisdoi. No experimental collision data are distributed with the software.

\section*{Declaration of competing interest}
The authors declare that they have no known competing financial interests or personal relationships that could have appeared to influence the work reported in this paper.

\section*{Declaration of generative AI and AI-assisted technologies in the writing process}
During preparation of this work, the authors used ChatGPT (OpenAI) to assist with consistency checks and the organisation of software and reproducibility descriptions. The authors reviewed and edited all generated material and take full responsibility for the content of the article.

\section*{Acknowledgements}
The authors thank Toby Satterthwaite for contributions to the wider \setanubis project and Oleg Brandt's supervision and coordination of the \setanubis project. The authors also acknowledge the developers and maintainers of the external scientific software cited in this article. 

\clearpage
\bibliographystyle{elsarticle-num}
\bibliography{bibs/anubis-standard-refs,bibs/main}

\clearpage
\appendix
\section{Installation, package distribution and optional dependencies}
\label{app:Installation}

\subsection{Release channels and supported Python versions}
\label{app:InstallReleaseChannels}
Release \setanubistag targets Python 3.10--3.13. The default distribution is a platform-independent, pure-Python wheel; the optional native Pythia interface is compiled only when explicitly requested. The release workflow builds one wheel/source-distribution pair, publishes it first to TestPyPI, downloads and verifies the wheel checksum, performs installation smoke tests, and only then promotes the identical artifacts to PyPI. Accordingly, TestPyPI is used for release rehearsal and integrity verification, while the supported end-user installation is obtained from PyPI.

Once the final tag has been published, an isolated installation is obtained with
\begin{lstlisting}[language=bash]
python3 -m venv .venv
source .venv/bin/activate
python -m pip install --upgrade pip
python -m pip install "SetAnubis==1.0.0"
python -c "import setanubis; print(setanubis.__version__)"
setanubis-pythia-smoke --out pythia_smoke_output
\end{lstlisting}
The last command validates pure-Python Pythia command generation and reports separately whether the optional compiled binding is available.

\subsection{Installation from a source checkout}
\label{app:InstallCheckout}
For an exact source checkout of the release described in this article, the tagged repository should be installed from its root directory. The editable installation keeps imports linked to the checkout and therefore does not require reinstallation after ordinary source changes.
\begin{lstlisting}[language=bash]
git clone --branch v1.0.0 --depth 1 https://github.com/SET-ANUBIS/set-anubis.git
cd set-anubis
python3 -m venv .venv
source .venv/bin/activate
python -m pip install --upgrade pip
python -m pip install -e ".[dev,docs,selection]"
\end{lstlisting}
For development on the current branch, the \texttt{{{-}{-}branch v1.0.0}} option can be omitted. The extra features are composable: \texttt{pythia} adds the build and HepMC Python requirements for the optional native interface, \texttt{selection} adds \texttt{pyhepmc}, \texttt{app} installs the interactive Dash applications, \texttt{docs} installs the Sphinx toolchain, and \texttt{dev} installs the test, build and static-analysis tools. The MadGraph Docker adapter is available in the base distribution. A base source installation can instead use \texttt{python -m pip install -e .}.

A release-scale verification can then be performed with:
\begin{lstlisting}[language=bash]
python -m pytest -q setanubis/tests --cov=SetAnubis --cov-config=pyproject.toml --cov-report=term-missing --cov-fail-under=58
python reproducibility/run_reproducibility.py --output-root reproducibility_outputs
\end{lstlisting}
For the supplied release snapshot these commands execute 247 tests and validate all five deterministic R1--R5 scenarios.

\subsection{Optional native Pythia8/HepMC3 binding}
\label{app:InstallPythia}
The PyPI wheel intentionally omits the native \texttt{pythia\_sim} extension. Running \pythia from Python requires compatible \pythia{8} and HepMC3 installations, a C++ compiler and \texttt{pybind11}. From a checkout, the project helper can build local copies and the extension can subsequently be requested during installation:
\begin{lstlisting}[language=bash]
./External_Integration/install.sh HepMC3 Pythia
EXTERNAL="\$PWD/External_Integration"
PYTHIA_PREFIX="\$EXTERNAL/Pythia/pythia8315"
HEPMC_PREFIX="\$EXTERNAL/HepMC3/hepmc3-install"
SETANUBIS_BUILD_PYTHIA=1 SETANUBIS_PYTHIA8_DIR="\$PYTHIA_PREFIX" SETANUBIS_HEPMC3_DIR="\$HEPMC_PREFIX" python -m pip install -e ".[pythia]"
setanubis-pythia-check
\end{lstlisting}
Existing system installations may be used by replacing the two prefix variables with their actual locations. The build fails explicitly when headers or shared libraries cannot be found; a standard Python-only installation remains functional for model handling, branching-ratio calculations, deterministic card construction and the non-native selection utilities.

\subsection{MadGraph, MARTY and system dependencies}
\label{app:InstallExternal}
\madgraph and \marty are external scientific applications and are not embedded in the Python wheel. The repository helper accepts the integration names \texttt{HepMC3}, \pythia, \madgraph and \marty; dependencies are resolved automatically, so requesting \pythia also installs HepMC3. For example,
\begin{lstlisting}[language=bash]
./External_Integration/install.sh MadGraph
./External_Integration/install.sh Marty
\end{lstlisting}
The local \madgraph helper currently pins \madversion{3.5.8} and targets Fedora/RHEL-like systems. An alternative adapter executes \madgraph in the project Docker image, which avoids a host installation but requires a functioning Docker daemon. The image is pulled automatically when the named container is created. Docker availability can be checked with:
\begin{lstlisting}[language=bash]
docker run --rm hello-world
\end{lstlisting}
For local builds, the complete event-generation stack requires a C/C++ compiler, CMake, Make and \texttt{gfortran}; package names depend on the operating system. Linux is the primary supported platform for the full external-tool workflow, with WSL documented for Windows hosts. Users who only inspect UFO models, prepare deterministic cards, evaluate supplied decay calculations or run the pure-Python release checks do not need to install all external generators.

\subsection{Packaged examples and installation diagnostics}
\label{app:InstallExamples}
The wheel includes the executable examples and compact validation resources used in Section~\ref{sec:ValidationAndExamples}. Representative post-installation checks are
\begin{lstlisting}[language=bash]
python -m SetAnubis.examples.ModelCore.example_setanubis_interface
python -m SetAnubis.examples.MadGraph.example_madgraph_interface
python -m SetAnubis.examples.Pythia.example_pythia_cmnd
python -m SetAnubis.examples.Selection.example_real_selection_trace_report \        --output-dir selection_trace_output
\end{lstlisting}
The \madgraph example serialises cards without starting the external program, and the \pythia command-generation example does not invoke the native generator. The selection example uses the packaged seven-event HNL benchmark and writes both JSON and standalone HTML traces, providing a compact end-to-end check of the event-representation, geometry and cutflow interfaces.

\section{QCD running}
\label{app:QCD}
This section documents the conventions and formulae implemented in the \texttt{QCDRunner} helper used in the framework's numerical results. It provides a three--loop, Next-to-Next-to Leading Order (NNLO), analytic approximation for the strong coupling \(\alpha_s(Q)\) at scale, Q; the NNLO running of quark masses in the \(\overline{\text{MS}}\) renormalisation scheme; and simple pole\(\leftrightarrow\)\(\overline{\text{MS}}\) conversions for \(b\) and \(t\) quarks. Electroweak effects and finite-order decoupling constants beyond continuity at thresholds are not included.

\subsection*{Notation}
Working in QCD with \(n_f\) active flavours, the following terms are defined:
\begin{align*}
L &\equiv \ln\!\frac{Q^2}{\Lambda_{n_f}^2}, &
\zeta_3 &\equiv 1.2020569031595942, &&
\end{align*}
where $\Lambda_{n_f}$ is the energy-scale in this scenario and $\zeta_3$ is the Riemann-Zeta function evaluated for $n=3$.

The one–, two– and three--loop coefficients of the QCD \(\beta\)-function and the quark mass anomalous dimension are (in the \(\overline{\text{MS}}\) scheme)

\begin{align*}
b_0 &= 11 - \frac{2}{3}n_f, \qquad b_1 = 51 - \frac{19}{3}n_f, \nonumber\\
b_2 &= 2857 - \frac{5033}{9}n_f + \frac{325}{27}n_f^2, \\
\gamma_0 &= 2, \qquad  \gamma_1 = \frac{101}{12} - \frac{5}{18}n_f, \nonumber\\
\gamma_2 &= \frac{1}{32}\!\left[1249
- \left(\frac{2216}{27}+\frac{160}{3}\zeta_3\right)n_f\right.\nonumber
\left.-\frac{140}{81}n_f^2\right]
\end{align*}

\subsection*{NNLO analytic approximation of \(\alpha_s(Q)\)}
The code evaluates \(\alpha_s(Q)\) with the standard three--loop expansion in \(1/L\) including \(\ln L\) terms as shown in Equation~\eqref{eq:alphas-analyticreal}.
For a given target value \(\alpha_s(Q_\star)\) at fixed \(n_f\), the corresponding energy scale, \(\Lambda_{n_f}\), is obtained by solving
${\alpha_s^{(n_f)}(Q_\star;\Lambda_{n_f})=\alpha_s(Q_\star)}$
via a bisection on \(\Lambda_{n_f}\).

\subsection*{Flavour thresholds and \(\Lambda\)-matching}
Let the ordered quark-mass thresholds be
\[
(m_u,\,m_d,\,m_s,\,m_c,\,m_b^{\text{(type)}},\,m_t^{\text{(type)}}).
\]
Here ``type'' denotes whether the threshold value uses a pole or running mass (configurable in the code). For any scale, \(Q\),
\[
n_f(Q) = N_f(\, m_q < Q \,),
\]
where $N_f$ is the number of flavours and $m_q$ is the quark mass.

Across each heavy–quark threshold \(Q=m_h\), \(\Lambda\) is matched by enforcing continuity of the coupling:
\begin{equation}
\alpha_s^{(n_f)}(m_h;\Lambda_{n_f}) \;=\; \alpha_s^{(n_f\pm1)}(m_h;\Lambda_{n_f\pm1}),
\label{eq:lambda-matching}
\end{equation}
which is solved for \(\Lambda_{n_f\pm1}\) using Equation~\eqref{eq:alphas-analyticreal}. Starting from \(\Lambda_5\) fixed by the input \(\alpha_s(M_Z)\), the code steps up/down in \(n_f\) as needed to evaluate \(\alpha_s(Q)\) at any scale.

\subsection*{Running quark masses (NNLO)}
For fixed \(n_f\), the quark mass obeys the differential equation
\({\mu\,\mathrm{d}\ln m/\mathrm{d}\mu = -\gamma_m(\alpha_s)}\),
with the NNLO solution implemented via the standard closed form
\begin{align}
m(Q_f) \;&=\; m(Q_i)\;
\frac{R\left(\alpha_s^{(n_f)}(Q_f)\right)}
     {R\left(\alpha_s^{(n_f)}(Q_i)\right)},
\qquad\\
R(\alpha) &\equiv
\Bigl(\tfrac{b_0 \alpha}{2\pi}\Bigr)^{2\frac{\gamma_0}{b_0}}
\Bigl[1 + B\,\tfrac{\alpha}{\pi}
     + \tfrac12\,C\,\Bigl(\tfrac{\alpha}{\pi}\Bigr)^{\!2}\Bigr],
\label{eq:mass-running}
\end{align}
with
\begin{align*}
B &= \frac{2\gamma_1}{b_0} - \frac{b_1\gamma_0}{b_0^2}, \\
C &= \left(\frac{2\gamma_1}{b_0} - \frac{b_1\gamma_0}{b_0^2}\right)^{\!2}
     + \frac{2\gamma_2}{b_0} - \frac{b_1\gamma_1}{b_0^2}
     - \frac{b_2\,\gamma_0}{(4 b_0)^2}
     + \frac{b_1^{2}\gamma_0}{2 b_0^{3}}.
\end{align*}
Running between arbitrary \(Q_i\!\to\!Q_f\) with thresholds in-between is performed piecewise:
for each interval where \(n_f\) is constant, apply Equation~\eqref{eq:mass-running}, then update \(Q_i\) to the next threshold and repeat until \(Q_f\) is reached.

\subsection*{Pole--\(\overline{\text{MS}}\) conversions used for \(b\) and \(t\)}
For the \(b\) quark (from running to pole). At the energy scale \({\mu=\bar m_b\equiv m_b^{\overline{\text{MS}}}(\bar m_b)}\),
\begin{align}
m_b^{\text{pole}} \;=\; \bar m_b
\biggl[
1 &+ \frac{\alpha_s(\bar m_b)}{\pi}
\biggl(
\frac{4}{3}+\nonumber\\
&\frac{\alpha_s(\bar m_b)}{\pi}
\Bigl(13.4434 - 4.1656+\nonumber\\
&\frac{1.3885}{\bar m_b}\,\sum_{q=u,d,s,c}\bar m_q(\bar m_q)\Bigr)
\biggr)
\biggr].
\label{eq:mb-pole}
\end{align}
The numerical constants reflect the fixed-order coefficients (including a finite light–quark mass correction proportional to \(\sum \bar m_q\)) used by the implementation.

For the \(t\) quark (from pole to running). The code solves for \(\bar m_t \equiv m_t^{\overline{\text{MS}}}(\bar m_t)\) from
\begin{align}
m_t^{\text{pole}} \;=\; \bar m_t\,
\biggl[
& 1 + \frac{\alpha_s(\mu)}{6\pi}
\biggl(
8 + \nonumber\\
& \frac{\alpha_s(\mu)}{\pi}
\Bigl(\frac{2053}{48} + 2\pi^2\bigl(1+\tfrac{\ln 2}{3}\bigr) - \zeta_3\Bigr)
\biggr)
\biggr],
\label{eq:mt-pole-to-ms}
\end{align}
iterating once with \(\mu=m_t^{\text{pole}}\) and then with \(\mu=\bar m_t\) to stabilise the result.

\subsection*{The algorithm (as implemented)}
\begin{enumerate}
\item Fix \(\Lambda_5\) from the input \(\alpha_s(M_Z)\) by inverting Equation~\eqref{eq:alphas-analyticreal}.
\item Determine the ordered thresholds\newline ${(m_u, m_d, m_s, m_c, m_b^{\text{(type)}}, m_t^{\text{(type)}})}$.
\item To evaluate \(\alpha_s(Q)\): step the flavour number from 5 to \(n_f(Q)\) using Equation~\eqref{eq:lambda-matching}, then use Equation~\eqref{eq:alphas-analyticreal} at \(n_f(Q)\).
\item To run a quark mass \(m(Q_i)\to m(Q_f)\): decompose the path into constant-\(n_f\) segments and apply Equation~\eqref{eq:mass-running} on each.
\item Convert \(m_b\) or \(m_t\) when needed using Equations~\eqref{eq:mb-pole} and~\eqref{eq:mt-pole-to-ms}.
\end{enumerate}

\subsection*{Practical settings and caveats}
\begin{itemize}
\item Inputs: \(\alpha_s(M_Z)\), \(M_Z\), \(m_t^{\text{pole}}\), \(\bar m_b\), and \(\bar m_{u,d,s,c}\) (all in GeV).
\item Threshold choice for \(b,t\) (pole or running) is configurable and only affects where \(n_f\) changes.
\item Matching uses \emph{continuity} of \(\alpha_s\) at \(Q=m_h\) without higher–order decoupling constants; this is adequate for this purpose.
\item All results are purely perturbative QCD (no electroweak corrections).
\end{itemize}

\end{document}